\documentclass[usenatbib]{mn2e}
\usepackage{tabularx,hhline,amsmath,amssymb}
\usepackage{graphics}
\usepackage[small]{subfigure}

\title[HMXB as a Star Formation Rate Indicator]{High Mass X-ray Binaries as
a Star Formation Rate Indicator in Distant Galaxies}
\author[H.-J. Grimm et al.]
       {H.-J. Grimm,$^1$,M.~Gilfanov,$^{1,2}$,R.~Sunyaev,$^{1,2}$\\
        $^1$Max-Planck-Institut f\"ur Astrophysik,
            Karl-Schwarzschild-Str. 1 85741 Garching b. M\"unchen, Germany\\
	$^2$Space Research Institute of Russian Academy of Sciences,
            Profsoyuznaya 84/32, 117810 Moscow, Russia}
\date{\today}

\pagerange{\pageref{firstpage}--\pageref{lastpage}}
\pubyear{200?}

\begin{document}

\maketitle

\begin{abstract}

Based on CHANDRA and ASCA observations of nearby starburst galaxies
and RXTE/ASM, ASCA, and MIR-KVANT/TTM studies of high mass X-ray
binary (HMXB) populations in the Milky Way and Magellanic Clouds, we
propose that the number and/or the collective X-ray luminosity of
HMXBs can be used to measure the star formation rate (SFR) of a
galaxy. We show that, within the accuracy of the presently available
data, a linear relation between HMXB number and star formation
rate exists. The relation between SFR and collective luminosity of
HMXBs is non-linear in the low SFR regime, $L_X\propto
\text{SFR}^{\sim 1.7}$, and becomes linear only for sufficiently
high star formation rate, SFR$\ga 4.5$ M$_{\odot}$ yr$^{-1}$ (for M$_{*}>
8$M$_{\odot}$). The non-linear $L_X-SFR$ dependence in the low SFR
limit is {\it not} related to non-linear SFR-dependent effects in the
population of HMXB sources. It is rather caused by the fact, that we
measure the collective luminosity of a population of discrete sources,
which might be dominated by the few brightest sources. Although more
subtle SFR-dependent effects are likely to exist, in the entire range
of SFRs the data are broadly consistent with the existence of a
universal luminosity function of HMXBs which can be roughly described
as a power law with a differential slope of $\sim 1.6$, a cutoff at
$L_X\sim{\rm few} \times 10^{40}$ erg s$^{-1}$ and a normalisation
proportional to the star formation rate.

We apply our results to (spatially unresolved) starburst galaxies
observed by CHANDRA at redshifts up to $z\sim 1.2$ in the Hubble Deep
Field North and show that the calibration of the collective luminosity
of HMXBs as a SFR indicator based on the local sample agrees well with
the SFR estimates obtained for these distant galaxies with
conventional methods. 

\end{abstract}

\section{Introduction}
\label{sec:intro}
X-ray observations open a new way to determine the star formation rate
(SFR) in young very distant galaxies. CHANDRA observations of actively
star forming galaxies in our vicinity, RXTE/ASM, ASCA, and
MIR-KVANT/TTM data about high-mass X-ray binary (HMXB) populations in
our Galaxy and the Magellanic Clouds provide a possibility to
calibrate the dependence of SFR on the X-ray luminosity of a galaxy
due to HMXBs. For nearby, spatially resolved galaxies for which
CHANDRA is able to resolve individual X-ray binaries we also have the
opportunity to calibrate the dependence of SFR on the total number of
HMXB sources.

In the absence of a bright AGN, the X-ray emission of a galaxy is
known to be dominated by the collective emission of its X-ray binary
populations (see e.g. \citet{fabbiano:94}). X-ray binaries,
conventionally divided into low and high mass X-ray binaries, consist
of a neutron star (NS) or a black hole (BH) accreting from a normal
companion star. To form a NS or BH the initial mass of the progenitor
star must exceed $\sim$ 8 M$_{\odot}$ \citep{verbunt:94}. The main
distinction between LMXBs and HMXBs is the mass of the optical
companion with a broad, thinly populated dividing region between $\sim
1-5$ M$_{\odot}$. This difference results in drastically different
evolution time-scales for low and high mass X-ray binaries and, hence,
different relations of their number and collective luminosity to the
instantaneous star formation activity and the stellar content of the
parent galaxy. In the case of a HMXB, having a high mass companion,
generally $M_{optical}\ga 10$ M$_{\odot}$ \citep{verbunt:94}, the
characteristic time-scale is at most the nuclear time-scale of the
optical companion which does not exceed $\sim 2\cdot 10^{7}$ years
whereas for a LMXB, generally $M_{optical}\la 1$ M$_{\odot}$, it is of
the order of $\sim 10^{10}$ years.

HMXBs were first recognised as short-living objects fed by the gas
supply of a massive star as a result of the discovery of Cen X-3 as an
X-ray pulsar by UHURU, in a binary system with an optical companion of
more than 17 M$_{\odot}$ \citep{schreier:72}, and the localisation and
mass estimation of the Cyg X-1 BH due to a soft/hard state
transition occurring simultaneously with a radio flare
\citep{tananbaum:72}, and following optical observations of a bright
massive counterpart \citep{bolton:72,lyutyi:73}. Dynamics of
interacting galaxies, e.g. Antennae, provide an additional upper limit
on the evolution and existence time-scale of HMXBs since the tidal
tails and wave patterns in which star formation is most vigorous are
very short-lived phenomena, of the order of a crossing time of
interacting galaxies \citep{toomre:72,eneev:73}.

The prompt evolution of HMXBs makes them a potentially good tracer of
the very recent star formation activity in a galaxy \citep{sunyaev:78}
whereas, due to slow evolution, LMXBs display no direct connection to
the present value of SFR. LMXBs rather are connected to the total
stellar content of a galaxy determined by the sequence of star
formation episodes experienced by a galaxy during its lifetime
\citep{ghosh:01,ptak:01,grimm:02}.

Several calibration methods are employed to obtain SFRs using UV,
FIR and radio flux from distant galaxies. Many of these methods rely
on a number of assumptions about the environment in the galaxy and
suffer from various uncertainties, e.g. the influence of dust, escape
fraction of photons or the shape of the initial mass function (IMF).
An additional and independent calibrator might therefore become a
useful method for the determination of SFR. Such a method, based on
the X-ray emission of a galaxy, might circumvent one of the main
sources of uncertainty of conventional SFR indicators -- absorption by
dust and gas. Indeed, galaxies are mostly transparent to X-rays above
about 2 keV, except for the densest parts of the most massive
molecular clouds.

The existence of various correlations between X-ray and
optical/far-infrared properties of galaxies has been noted and studied
in the past. Based on Einstein observations of normal and starburst
galaxies from the IRAS Bright Galaxy Sample, \citet{griffiths:90} and
\citet{david:92} found correlations between the soft X-ray luminosity
of a galaxy and its far-infrared and blue luminosity. Due to the
limited energy range (0.5--3 keV) of the Einstein observatory data one
of the main obstacles in quantifying and interpreting these
correlations was proper accounting for absorption effects and
intrinsic spectra of the galaxies which resulted in considerable
spread in the derived power law indices of the X-ray -- FIR
correlations, $\sim 0.7-1.0$. Moreover, supernova remnants are bright
in the soft band of the Einstein observatory. CHANDRA, however, is
able to distinguish SNRs from other sources due to its sensitivity to
harder X-rays. Although the X-ray data were not sufficient to
discriminate between contributions of different classes of X-ray
sources, \citet{david:92} suggested that the existence of such
correlations could be understood with a two component model for X-ray
and far-infrared emission from spiral galaxies, consisting of old and
young populations of the objects having different relations to the
current star formation activity in a galaxy. The uncertainty related
to absorption effects was recently eliminated by \citet{ranalli:02},
who extended these studies to the harder energy band of 2--10 keV
based on BeppoSAX and ASCA data. In particular, they found a
linear correlation between total X-ray luminosity of a galaxy and both
radio and far-infrared luminosities and suggested that the X-ray
emission might be directly related to the current star formation rate
in a galaxy and that such a relation might also hold at higher
redshifts.

\begin{table*}
\caption{The primary sample of local galaxies used to study the
luminosity function of HMXB sources.}
\begin{tabular}{|l|cc|r|r|r|r|r|r|r|}
\hline
Source  & Hubble & distance$^{(b)}$ & SFR$^{(c)}$ & M$_{dynamical}$ &
Ref.$^{(d)}$ & SFR/M & $N(L>$~~~~~~~& $L_{X,total}$ & Ref.$^{(e)}$ \\
  & type$^{(a)}$   & [Mpc] & [M$_{\odot}$ yr$^{-1}$] & [$10^{10}$M$_{\odot}$] & &
[$10^{-10}$yr$^{-1}$]&$2\cdot10^{38}$erg s$^{-1}$) & [$10^{39}$ erg s$^{-1}$] &\\
\hline
N3256            & Sb(s) pec & 35.0 & 44.0 &  5.0 & (i)   &  8.8  & 12 & 128& (1) \\
Antennae         & Sc pec    & 19.3 &  7.1 &  8.0 & (i)   &  0.9  & 27 & 49 & (2) \\
M 100            & Sc(s)     & 20.4 &  4.8 & 24.0 & (ii)  &  0.2  &  5 & 10 & (3) \\
M 51             & Sbc(s)    &  7.5 &  3.9 & 15.0 & (iii) &  0.26 & 15 & 16 & (4) \\
M 82             & Amorph    &  5.7 &  3.6 &  1.0 & (iv)  &  3.6  & 12 & 23 & (5) \\
M 83             & SBc(s)    &  3.8 &  2.6 & 15.4 & (v)   &  0.17 &  2 &0.14& (6) \\
N4579$^{(f)}$    & Sab(s)    & 23.5 &  2.5 &  -   &  -    &   -   &  5 & 26 & (7) \\
M 74             & Sc(s)     & 12.0 &  2.2 & 14.3 & (vi)  &  0.15 &  8 & 14 & (8) \\
Circinus$^{(g)}$ & Sb        &  3.7 &  1.5 &  2.2 & (v)   &  0.73 &  6 & 5  & (9) \\
N4736            & RSab(s)   &  4.5 &  1.1 &  7.0 & (v)   &  0.16 &  4 & 4  & (6) \\
N4490            & Scd pec   &  8.6 &  1.0 &  2.3 & (vii) &  0.43 &  2 & 1.2& (10)\\
N1569            & Sm        &  2.1 &  0.17&  0.03& (viii)&  5.6  &  0 & 0.2& (11)\\
SMC              & Im        &  0.06&  0.15&  0.2 & (ix)  &  0.75 &  1 & 0.4& (12)\\
\hline							      	      	   
Milky Way        & SAB(rs)bc &  -   &  0.25& 54   & (x)   &  0.005&  0 & 0.2& (13)\\
\hline
\end{tabular}
\label{tab:gal1}
\flushleft
$^{(a)}$from \citet{sandage:80}\\

$^{(b)}$assuming $H_0 = 70$ km s$^{-1}$ Mpc$^{-1}$ and using velocities
from \citet{sandage:80}.\\

$^{(c)}$ adopted SFR value -- from the last column of Table \ref{tab:flux}\\

$^{(d)}$References for masses: (vi) \citet{sharina:96}, (iv)
\citet{sofue:92}, (i) \citet{lipari:00}, (ii) \citet{persic:88},
(v) \citet{huchtmeier:88}, (iii) \citet{kuno:97}, (vii) \citet{sage:93}
(ix) \citet{feitzinger:80}, (x) \citet{wilkinson:99}, (viii) \citet{reakes:80}\\

$^{e}$References for X-ray observations: (1) \citet{lira:02},
(2) \citet{zezas:02}, (3) \citet{kaaret:01}, (4) \citet{terashima:02},
(5) \citet{griffiths:00}, (6) \citet{soria:02}, (7) \citet{eracleous:02},
(8) \citet{soria:02b}, (9) \citet{smith:01}, (10) \citet{roberts:02},
(11) \citet{martin:02}, (12) \citet{yokogawa:00}, 
(13) \citet{grimm:02}\\

$^{(f)}$We were not able to obtain a mass value for this source, but
according to the rotation curve it is not more massive than the
Milky Way \citep{gonzalez:96}.\\

$^{(g)}$Hubble type and velocity taken from \citet{devaucouleurs:91}.\\
\end{table*}

\begin{table*}
\caption{The secondary sample of local galaxies used to complement the
primary sample in the analysis of the $L_X$-SFR relation.}
\begin{center}
\begin{tabular}{|l|cc|r|r|r|r|r|r|}
\hline
Source  & Hubble & distance$^{(b)}$ & SFR$^{(c)}$ & M$_{dynamical}$ & Ref.$^{(d)}$ &
SFR/M & $L_{X,total}$ & Ref.$^{(e)}$ \\
  & type$^{(a)}$   & [Mpc] & [M$_{\odot}$ yr$^{-1}$] &
[$10^{10}$M$_{\odot}$] & & [$10^{-10}$yr$^{-1}$]& [$10^{39}$ erg s$^{-1}$] &\\
\hline
N3690            & Spec      & 44.3 & 40.0 &  -   &  -   &   -    &220  & (1)\\
N7252            & merger    & 68.3 &  7.7 &  4.0 & (i)  &  1.9   &94.6 & (2)\\
N253             & Sc(s)     &  4.2 &  4.0 &  7.3 & (ii) &  0.55  & 5.1 & (3)\\
N4945            & Sc        &  3.9 &  3.1 &  9.3 & (ii) &  0.41  & 8.9 & (4)\\
N3310            & Sbc(r)pec & 15.3 &  2.2 &  2.0 & (iii)&  1.1   &49.0 & (1)\\
N891             & Sb        & 11.1 &  2.1 & 24.0 & (iv) &  0.09  &31.0 & (1)\\
N3628            & Sbc       & 10.3 &  1.6 & 16.0 & (v)  &  0.1   &13.9 & (5)\\
IC342$^{(g)}$    & Scd       &  3.5 &  0.48& 11.8 & (ii) &  0.04  & 0.9 & (1)\\
LMC$^{(h)}$      &           &  0.05&  0.25&  0.5 & (vi) &  0.5   & 0.34& (6)\\
\hline
\end{tabular}
\end{center}
\label{tab:gal2}
\flushleft
$^{(a)}$from \citet{sandage:80}.\\

$^{(b)}$assuming $H_0 = 70$ km s$^{-1}$ Mpc$^{-1}$ and using velocities
from \citet{sandage:80}.\\

$^{(c)}$ adopted SFR value -- from the last column of Table
\ref{tab:flux}, or computed from data of \citet{condon:90} and
\citet{moshir:93}\\

$^{(d)}$References for masses: (i) \citet{lipari:00},
(ii) \citet{huchtmeier:88}, (iii) \citet{galletta:82},
(iv) \citet{bahcall:83}, (v) \citet{sage:93},
(vi) \citet{feitzinger:80}\\

$^{e}$References for X-ray observations: (1) \citet{ueda:01}
(2) \citet{awaki:02}, (3) \citet{rephaeli:02},
(4) \citet{schurch:02}, (5) \citet{strickland:01},
(6) \citet{grimm:02}\\

$^{(f)}$We were not able to obtain a mass value for this source, but
the very high SFR ensures for any reasonable mass domination by
HMXBs.\\

$^{(g)}$Velocity and Hubble type from \citet{karachentsev:97}.\\

$^{(h)}$For LMC data see discussion in the text.\\
\end{table*}

The main surprise of the study presented here is that in the low SFR
regime the relation between SFR and collective luminosity of HMXBs
is non-linear, $L_X\propto {\rm SFR}^{\sim 1.7}$, and becomes
linear only for sufficiently high star formation rates, when the total
number of HMXB sources becomes sufficiently large. The non-linear
$L_X-SFR$ dependence is caused by the fact that we measure the
collective luminosity, that strongly depends on the brightest sources,
of a population of discrete sources. We give a qualitative and
approximate analytic treatment of this (purely statistical) effect
below and will discuss it in more detail in a separate paper
\citep{gilfanov:02}.

There are, however, two main obstacles to use the X-ray luminosity of
a galaxy as a SFR indicator. Firstly, if an active nucleus (AGN) is
present in a galaxy it can easily outshine HMXBs in X-rays. In
principle, the presence of an AGN component might be identified and,
in some cases separated, due to different X-ray spectra of an AGN and
X-ray binaries, provided a sufficiently broad band energy
coverage. Secondly, there is the dichotomy into LMXBs and HMXBs which
both have somewhat similar spectra that also could probably be
distinguished provided sufficiently broad band coverage and
sufficient signal-to-noise ratio. To estimate the SFR one is
interested only in the luminosity of HMXBs, therefore the LMXB
contribution needs to be subtracted. This could, in principle, be done
based on an estimate of the stellar mass of a galaxy. The results of a
study of the X-ray binary population of our Milky Way \citep{grimm:02}
and knowledge of the Galactic SFR allow to estimate at which point the
emission of HMXBs dominates the emission of a galaxy. This obviously
depends on the ratio of SFR to stellar mass of a galaxy. We found,
that roughly at a ratio of $\sim 0.05$ M$_{\odot}$ yr$^{-1}$ per $10^{10}$
M$_{\odot}$ of total dynamical mass, or $\sim 0.5$ M$_{\odot}$ yr$^{-1}$ per
$10^{10}$ M$_{\odot}$ of stellar mass, the emission of HMXB sources
begins to dominate the X-ray emission of a galaxy (where SFR value
refers to a formation rate of stars more massive than $\sim 5$
M$_\odot$). It should be  emphasised, however, that even in the worst
case the X-ray luminosity based SFR estimate should be able to provide
an upper limit on the ongoing star formation activity in a galaxy.

Future observations with present, CHANDRA and XMM, and upcoming X-ray
missions, Astro-E and especially Constellation-X and XEUS, the last
having 1 arc sec angular resolution and a 100 times larger effective
area than CHANDRA, will permit to get information about the SFR of
galaxies from X-rays even at high redshifts. We know from optical and
radio data that the SFR was much higher in galaxies at z $\sim$ 2--5
\citep{madau:00}. Therefore we could expect that in these galaxies the
contribution of HMXBs was strongly exceeding the contribution of
LMXBs.

\section{Sample of galaxies}

The list of galaxies used in the following analysis is given in
Table \ref{tab:gal1} and Table \ref{tab:gal2} along with their Hubble
type, distances and other relevant parameters. 

As our primary sample of local galaxies, used to study the HMXB
luminosity function and to calibrate the $L_X$--SFR relation,  we
chose a number of nearby late-type/starburst galaxies observed by
CHANDRA.  We based our selection primarily on two criteria. Firstly,
we selected galaxies that can be spatially resolved by CHANDRA
sufficiently well so that the contribution of a central AGN can be
discriminated and the luminosity functions of the compact sources can
be constructed without severe confusion effects. We should note,
however, that for the most distant galaxies from our primary sample
(e.g. NGC 3256), the probability of source confusion might become
non-negligible. Secondly, we limited our sample to galaxies known to
have high star formation rates, so that the population of X-ray
binaries is dominated by HMXBs and the contribution of low mass X-ray
binaries can be safely ignored (see subsection \ref{sec:lmxb} for more
detailed discussion).

In order to probe the HMXB luminosity function in the low SFR regime,
we used the results of the X-ray binary population study in the Milky
Way by \citet{grimm:02}, based on RXTE/ASM observations and the
luminosity function of high mass X-ray binaries in the Small
Magellanic Cloud obtained by ASCA \citep{yokogawa:00}.

The galaxies from our primary sample are listed in Table \ref{tab:gal1}. 

In addition, in order to increase the local sample, we selected galaxies
observed by other X-ray missions, mainly ASCA, for which no luminosity
function is available, but a total flux measurement. The selection was
based on the requirement that no AGN-related activity had been
detected and the SFR to total mass ratio is sufficiently high to
neglect the LMXB contribution. These galaxies were used to complement
the primary sample in the analysis of the $L_X$--SFR relation. They
are listed in Table \ref{tab:gal2}. 

Finally, in order to study the $L_X$--SFR relation in distant galaxies
at redshifts of $z\sim 0.2-1.3$ we used a number of galaxies detected
by CHANDRA in the Hubble Deep Field North, see Table \ref{tab:hdfn}.
The selection criteria are similar to those applied to the local
sample and are described in more detail in Sec. \ref{sec:hdfn}.

\begin{figure*}
  \resizebox{0.49\hsize}{!}{\includegraphics{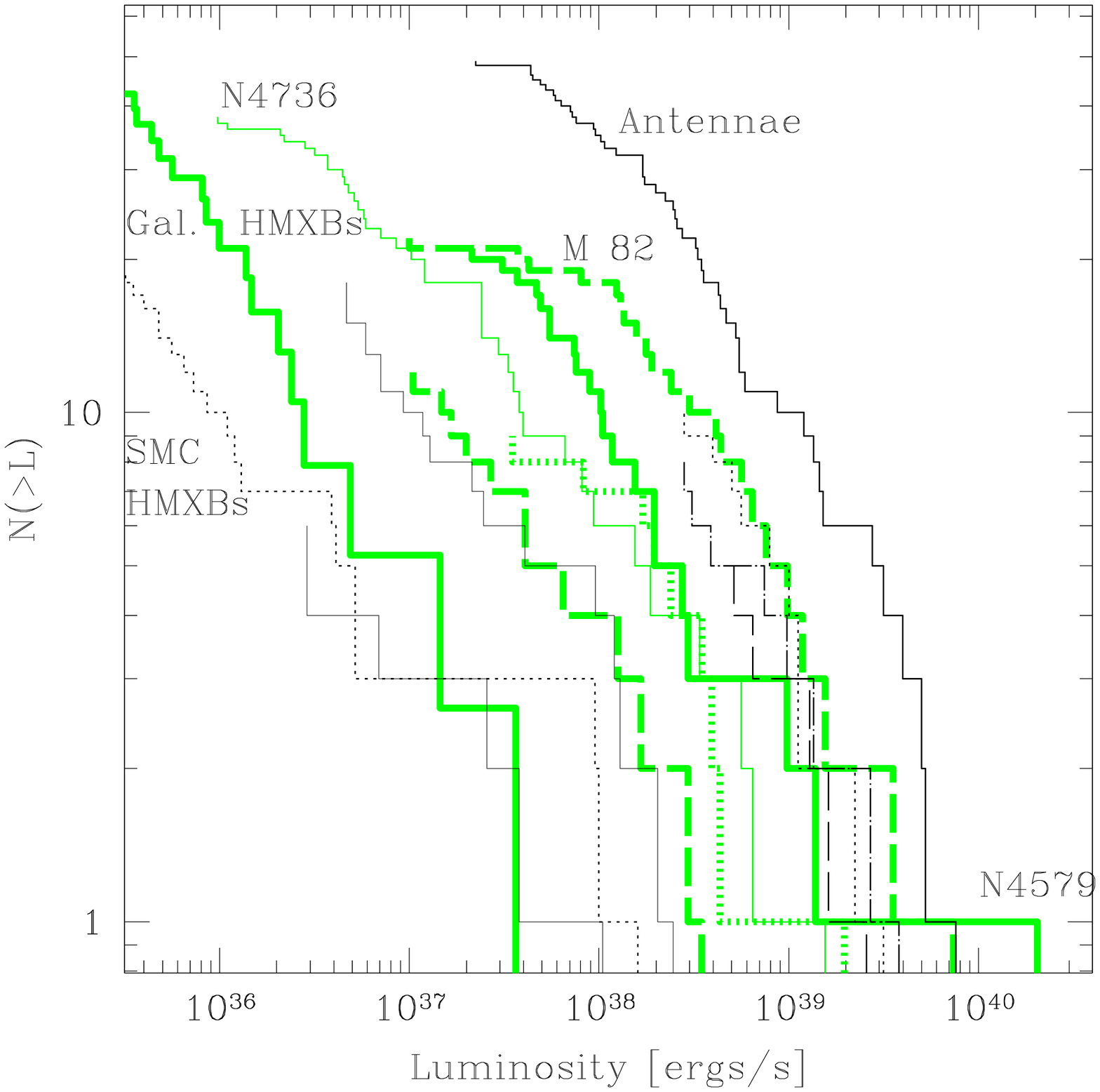}}
  \resizebox{0.49\hsize}{!}{\includegraphics{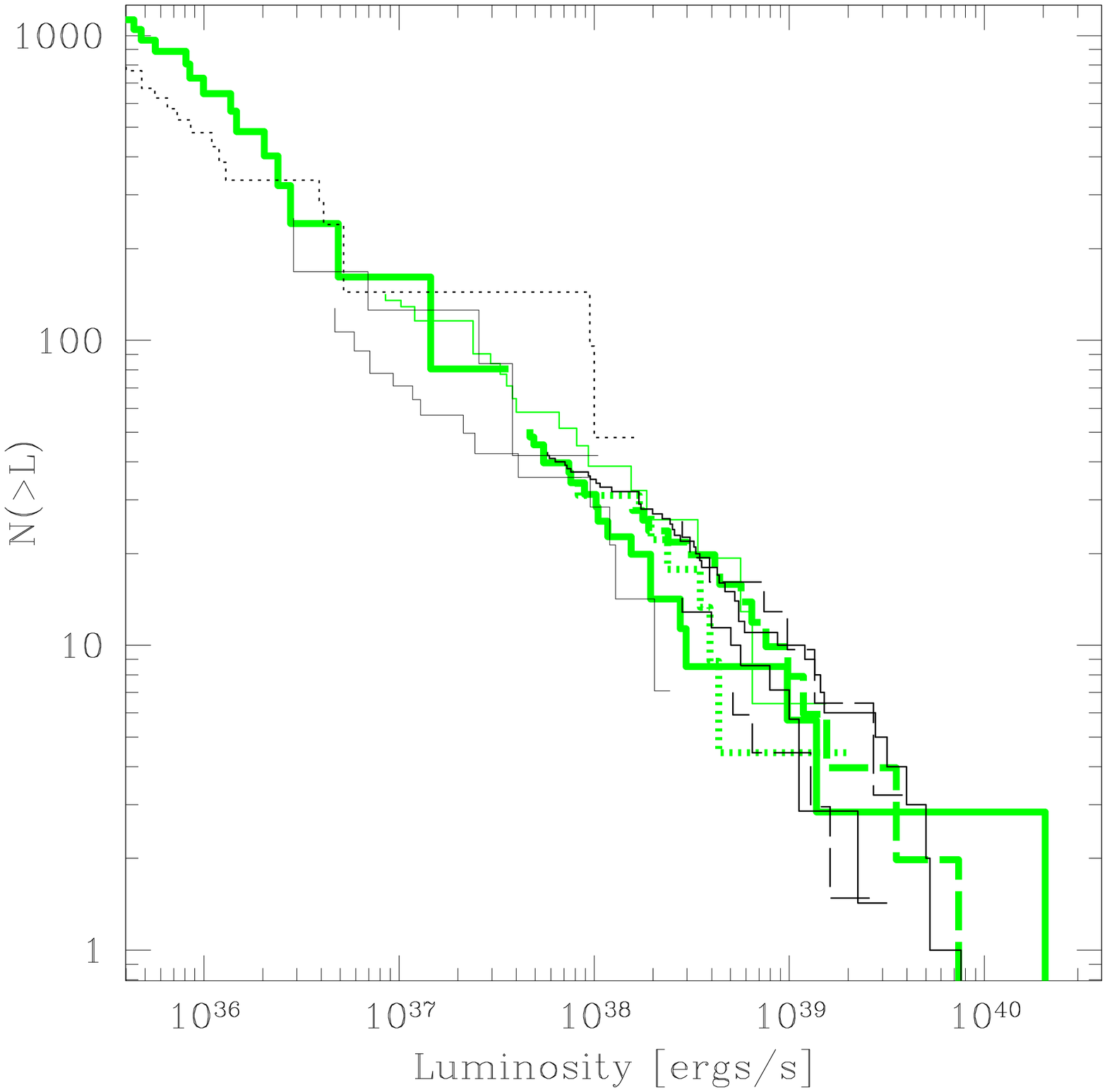}}
  \caption{{\em Left:} The luminosity functions of compact X-ray
  sources in nearby galaxies from the primary sample obtained by
  CHANDRA and listed in Table \ref{tab:gal1}. The
  luminosity functions are plotted assuming the distances from Table
  \ref{tab:gal1}. {\em Right:} The luminosity functions for the same
  galaxies scaled by the ratio of their star formation rate to the SFR
  of Antennae. The luminosity functions in the right panel are plotted
  only above their corresponding completeness limits. It is clear that
  despite large differences in the SFRs (by a factor of $\sim 40-50$)
  the scaled luminosity functions occupy only a narrow band in the
  $N(>L)-L$ plane.}
\label{fig:lfs}
\end{figure*}

\subsection{Distances}

To estimate X-ray luminosity and star formation rate, which is also
based on flux measurements in different spectral bands, and compare
these values for different galaxies it is necessary to have a
consistent set of distances. For the galaxies from our sample, given
in Tables \ref{tab:gal1} and \ref{tab:gal2} cosmological effects are
not important. The distances were calculated using velocities from
\citet{sandage:80} corrected to the centre of mass of the Local Group
and assuming a Hubble constant value of $H_0 = 70$ km s$^{-1}$ Mpc$^{-1}$. The
distances are listed in Table \ref{tab:gal1} and Table
\ref{tab:gal2}. Note that these distances might differ from the values
used in the original publications on the X-ray luminosity functions
and SFRs.

\begin{table*}
\caption{The star formation rates for the galaxies from the local sample,
measured by different SFR indicators.}
\begin{tabular}{|l|cccc|r|cccc|r|}
\hline
Source& \multicolumn{4}{|c|}{Fluxes}& &\multicolumn{4}{|c|}{SFRs}&\\
 & UV$^{1}$ & H$\alpha$$^{2}$ & FIR$^{3}$ & radio$^{4}$ & Reference&
UV & H$\alpha$ & FIR & radio &
adopted SFR \\ 
\hline
N3256    &        &  0.33  &   4.68 &        & (a)  &     &  5.3 & 31.0 &    & \\
         &        &        &   7.1  &        & (b)  &     &      & 47.0 &    & \\
         &        &        &   8.2  &        & (c)  &     &      & 54.0 &    & 44\\
\hline
N4038/9  &        &  1.62  &        &        & (d)  &     &  7.9 &      &    & \\
(Antennae)& 3.22  &  1.36  &   4.50 & 10.90  & (e)  & 9.2 &  6.7 &  9.0 & 9.1& \\
         &        &        &   2.30 &        & (f)  &     &      &  4.6 &    & 7.1\\
\hline
M 100    &        &  0.81  &        &        & (d)  &     &  4.5 &      &    & \\
         & 3.07   &  0.72  &   3.36 &        & (e)  & 9.8 &  3.9 &  7.5 &    & \\
         &        &        &   1.48 &        & (f)  &     &      &  3.3 &    & 4.8\\
\hline
M 51     & 15.4   &  3.45  &  14.7  &  8.62  & (e)  & 6.6 &  2.6 &  4.5 & 1.1& \\
         &        &  4.68  &        &        & (d)  &     &  3.5 &      &    & \\
         &        &  2.81  &        &        & (g)  &     &  2.1 &      &    & 3.9\\
\hline
M 82     &        &  6.17  &  52.0  &        & (h)  &     &  2.6 &  9.1 &    & \\
         &        &  9.12  &        &        & (d)  &     &  3.9 &      &    & \\
         & 1.46   &  9.98  & 112.0  & 76.70  & (e)  & 0.4 &  4.3 & 19.6 & 5.6& \\
         &        &        &  53.0  &        & (f)  &     &      &  9.2 &    & 3.6\\
\hline
M 83     &        & 13.50  &        &        & (i)  &     &  2.6 &      &    & \\
         &        &  0.45  &        &        & (j)  &     &  0.1 &      &    & \\
         & 32.4   & 12.20  &  34.2  &        & (e)  & 3.6 &  2.3 &  2.7 &    & 2.6\\
\hline
N4579    &        &  0.36  &        &        & (i)  &     &  2.6 &      &    & \\
         &        &  0.32  &        &        & (d)  &     &  2.4 &      &    & \\
         &        &        &   0.43 &        & (f)  &     &      &  1.3 &    & 2.5\\
\hline
M 74     &        &  1.23  &        &        & (d)  &     &  2.3 &      &    & \\
         & 6.85   &  1.25  &   2.92 &        & (e)  & 7.6 &  2.4 &  2.3 &    & \\
         &        &  1.51  &        &        & (g)  &     &  2.9 &      &    & \\
         &        &        &   1.59 &        & (f)  &     &      &  1.2 &    & 2.2\\
\hline
Circinus &        &        &  22.3  &        & (c)  &     &      &  1.6 &    & \\
         &        &  9.5   &  16.5  &        & (k)  &     &  1.7 &  1.2 &    & 1.5\\
\hline
N4736    &        &  5.37  &        &        & (d)  &     &  1.6 &      &    & \\
         &        &  5.37  &        &        & (i)  &     &  1.6 &      &    & \\
         & 6.49   &  2.10  &   6.78 &  5.80  & (e)  & 1.1 &  0.6 &  0.8 & 0.3& 1.1\\
\hline
N4490	 &        &  1.10  &   4.42 &        & (l)  &     & 1.1  &  1.8 &    & \\
	 &        &        &   2.31 &        & (m)  &     &      &  0.9 &    & \\
	 &        &        &        &85$^{a}$& (n)  &     &      &      & 1.0& 1.0\\
\hline
\hline
N253     & 16.1   &  6.06  & 100.0  &  75.4  & (e)  & 2.2 &  1.4 &  9.5 & 3.0& \\
         &        &  6.46  &        &        & (d)  &     &  1.5 &      &    & \\
         &        &  6.38  &        &        & (o)  &     &  1.5 &      &    & \\
         &        &        &  68.7  &        & (c)  &     &      &  6.5 &    & \\
         &        &        &  70.1  &        & (f)  &     &      &  6.7 &    & 4.0\\
\hline
N1569    &        &  2.29  &        &        & (d)  &     &  0.15&      &    & \\
         &        &  3.14  &        &        & (o)  &     &  0.2 &      &    & \\
         &        &  2.95  &        &        & (p)  &     &  0.19&      &    & \\
         &        &        &   4.59 &        & (q)  &     &      &  0.12&    & 0.17\\
\hline
N3628    &        &  0.32  &        &        & (p)  &     &  0.4 &      &    & \\
         &        &        &   3.36 &        & (f)  &     &      &  1.9 &    & \\
         &        &        &   3.12 &        & (r)  &     &      &  1.8 &    & \\
         &        &        &   4.17 &        & (k)  &     &      &  2.4 &    & 1.6\\
\hline
N4945    &        &  4.43  &  55.8  &        & (k)  &     &  0.8 &  4.6 &    & \\
         &        &        &  46.2  &        & (c)  &     &      &  3.8 &    & \\
         &        &        &  37.0  &        & (r)  &     &      &  3.0 &    & 3.1\\
\hline
N7252    &        &        &   0.30 &        & (s)  &     &      &  7.6 &    & \\
         &        &        &   0.31 &        & (t)  &     &      &  7.8 &    & 7.7\\
\hline
\end{tabular}
\label{tab:flux}
\flushleft
Flux units: $^{1}$ -- $10^{-25}$ erg s$^{-1}$ cm$^{-2}$ Hz$^{-1}$; 
$^{2}$ -- $10^{-11}$ erg s$^{-1}$ cm$^{-2}$; 
$^{3}$ -- $10^{-9}$ erg s$^{-1}$ cm$^{-2}$;
$^{4}$ -- $10^{-25}$ erg s$^{-1}$ cm$^{-2}$ Hz$^{-1}$\\
References: (a) \citet{buat:02}, (b) \citet{lipari:00}, (c)
\citet{negishi:01}, (d) \citet{young:96}, (e) \citet{bell:01},
(f) \citet{david:92}, (g) \citet{hoopes:01}, (h) \citet{armus:90},
(i) \citet{roussel:01}, (j) \citet{rosa-gonzales:02},
(k) \citet{lehnert:96}, (l) \citet{thronson:89},
(m) \citet{viallefond:80}, (n) \citet{fabbiano:88},
(o) \citet{rownd:99}, (p) \citet{kennicutt:94}, (q) \citet{israel:88},
(r) \citet{rice:88}, (s) \citet{liu:95}, (t) \citet{georgakakis:00}\\
$^{a}$ non-thermal flux, SFR conversion with formula \ref{eq:ra2}
\end{table*}

\subsection{X-ray luminosity functions}
\label{sec:lumf}

For the X-ray luminosity functions we used published results of
Chandra observations of late-type/starburst galaxies. References to
the original publications are given in Table \ref{tab:gal1} and Table
\ref{tab:gal2}. The luminosities were rescaled to the distances
described in the previous subsection. Note that, due to this
correction, the total X-ray luminosities and luminosities of the
brightest sources might differ from the numbers given in the original
publications. The complete set of luminosity functions for all objects
from the primary sample (Table \ref{tab:gal1}) is plotted in
Fig. \ref{fig:lfs} (left panel).

One of the most serious issues important for the following analysis
is the completeness level of the luminosity functions which is
obviously different for different galaxies, due to different exposure
times and distances. In those cases when the completeness luminosity
was not given in the original publication, we used a conservative
estimate based on the luminosity at which the luminosity function
starts to flatten.

Due to the relatively small field of view of Chandra and sufficiently
high concentration of X-ray binaries in the central parts of the
galaxies the contribution of foreground and background objects can be
neglected for the purpose of our analysis (e.g. M 51
\citep{terashima:02}, M 83 \citep{soria:02}). Two of the galaxies
in our sample -- Circinus and NGC 3256 -- are located at a Galactic
latitude of $|b_{II}| < 20\degr$. In these cases the contribution of
foreground optical stars in the Galaxy that are bright in X-rays can
be discriminated based on the softness of their spectra. Extrapolating
the luminosity function of X-ray binaries in the Milky Way
\citep{grimm:02} the probability can be estimated of occurence of a
foreground source due to an unknown Galactic X-ray binary with a flux
exceeding the sensitivity limit of the corresponding Chandra
observations. For the Chandra field of view this probability is less
than $\sim 10^{-3}$ and therefore can be neglected.

The luminosities of the compact sources were derived in the original
publications in slightly different energy bands, under different
assumptions about spectral shape, and with different absorption column
densities. Although all these assumptions affect the luminosity
estimates, the resulting uncertainty is significantly smaller than
those due to distance uncertainty and uncertainties in the star
formation rate estimates. Moreover, in many cases, due to insufficient
statistics of the data an attempt to do corrections for these effects
could result in additional uncertainties, larger than those due to a
small difference in e.g. energy bands. Therefore we make no
attempt to correct for these differences. It should be mentioned
however, that the most serious effect, up to a factor of a few in
luminosity might be connected with intrinsic absorption for the
sources embedded in dense starforming regions
\citep{zezas:02}. Appropriately accounting for this requires
information about these sources, which is presently not available.

All the luminosity functions with exception of the Milky Way are
``snapshots'' of the duration of several tens of kiloseconds. On the
other hand, similar to the Milky Way, compact sources in other
galaxies are known to be variable. E.g. NGC 3628 is dominated by a
single source, that is known to vary by about a factor of 30
\citep{strickland:01}. This may affect the shape of the individual
luminosity functions. It should not however affect our conclusions,
since in the high SFR regime they are based on the average properties
of sufficiently many galaxies. As for the low SFR regime, the Milky
Way data are an average of the RXTE/ASM observations over four years
therefore the contribution of ``standard'' Galactic transient sources
is averaged out.

\subsection{Star formation rate estimates}

One of the main uncertainties involved is related to the SFR
estimates. The conventional SFR indicators rely on a number of
assumptions regarding the environment in a galaxy, such as dust
content of the galaxy, or the shape of the initial mass function
(IMF). Although comparative analysis of different star formation
indicators is far beyond the scope of this paper, in order to roughly
assess the amplitude of the uncertainties in the SFR estimates we
compared results of different star formation indicators for each
galaxy from our sample with special attention given to the galaxies
from the primary sample. For all galaxies from the primary sample we
found at least 3 different measurements of star formation indicators
in the literature, namely UV, H$\alpha$, FIR or thermal radio
emission flux. The data along with the corresponding references are
listed in Table \ref{tab:flux}.

In order to convert the flux measurements to star formation rates we
use the result of an empirical crosscalibration of star formation rate
indicators by \citet{rosa-gonzales:02}. The calibration is based on
the canonical formulae by \citet{kennicutt:98} and takes into account
dust/extinction effects. We used the following flux--SFR relations:
\begin{eqnarray}
\label{eq:ha}
  &SFR_{H\alpha} = 1.1 \cdot 10^{-41} \cdot L_{H\alpha}
  \text{[erg s$^{-1}$]}\\
\label{eq:uv}
  &SFR_{UV} = 6.4 \cdot 10^{-28} \cdot L_{UV} \text{[erg s$^{-1}$
  Hz$^{-1}$]}\\
\label{eq:ir}
  &SFR_{FIR} = 4.5 \cdot 10^{-44} \cdot L_{FIR} \text{[erg s$^{-1}$]}\\
\label{eq:ra}
  &SFR_{radio} = 1.82 \cdot 10^{-28} \cdot
  \nu_{\text GHz}^{0.1}\cdot L_{\nu} \text{[erg s$^{-1}$ Hz$^{-1}$]}
\end{eqnarray}
The last relation is from \citet{condon:92} and applies only to the
thermal radio emission, originating, presumably, in hot gas in HII
regions associated with star formation (as we used thermal 1.4 GHz
flux estimates from \citet{bell:01}).

The above relations refer to the SFR for stars more massive than
$\sim 5$ M$_{\odot}$. The total star formation rate, including low
mass stars, could theoretically be obtained by extrapolating these
numbers assuming an initial mass function. Obviously, such a
correction would rely on the assumption that the IMF does not depend
on the initial conditions in a galaxy and would involve a significant
additional uncertainty. On the other hand, this correction is not
needed for our study as the binary X-ray sources harbour a compact
object -- a NS or a BH -- which according to the
modern picture of stellar evolution can evolve only from stars with
initial masses exceeding $\sim 8$ M$_{\odot}$. The SFR correction from
$M>5$ M$_{\odot}$ to $M>8$ M$_{\odot}$ is relatively small ($\sim$
20 per cent) and, most importantly, due to the similarity of the IMFs for
large masses it is significantly less subject to the uncertainty due
to poor knowledge of the slope of the IMF. Thus, for the purpose of
our study it is entirely sufficient to use the relations
(\ref{eq:ha})--(\ref{eq:ra}) without an additional correction. In the
following, the term SFR refers to the star formation rate of stars
more massive than $\sim 5$ M$_{\odot}$.

Since the relations (\ref{eq:ha})--(\ref{eq:ra}) are based on the
average properties of star forming galaxies there is considerable
scatter in the SFR estimates of a galaxy obtained using different
indicators (Table \ref{tab:flux}). On the other hand, the SFR
estimates based on different measurements of the same indicator are
generally in a good agreement with each other. A detailed study, which
SFR indicator is most appropriate for a given galaxy is beyond the
scope of this paper. Therefore, we relied on the fact that for all
galaxies from our primary sample there are more than 3 measurements
for different indicators. For each galaxy we disregarded the estimates
significantly deviating from the majority of other indicators, and
averaged the latter. The final values of the star formation rates we
used in the following analysis are summarised in the last column of
Table \ref{tab:flux}.

\begin{figure}
  \resizebox{\hsize}{!}{\includegraphics{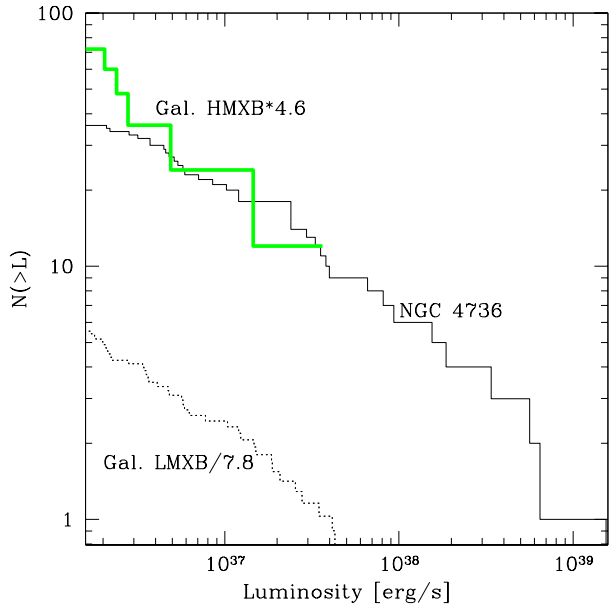}}
  \caption{Contributions of LMXBs and HMXBs to the observed luminosity
  function for NGC 4736 (thin solid histogram), having smallest SFR to
  total mass ratio in the primary sample. The upper thick grey
  histogram corresponds to the contribution of HMXBs scaled from the
  Milky Way HMXB luminosity function by the ratio of the SFRs. The
  lower dotted histogram is the Galactic LMXB luminosity function
  scaled by the ratio of the total masses. Total masses and SFRs are
  given in Table \ref{tab:gal1}.}
\label{fig:lmxb}
\end{figure}

\subsection{Contribution of a central AGN}

As mentioned in Sec. \ref{sec:intro} the emission of a central AGN
can easily outshine the contribution of X-ray binaries. However, due
to the excellent angular resolution of CHANDRA it is possible to
exclude any contribution from the central AGN in nearby galaxies. In
our primary sample a central AGN is present in the Circinus galaxy and
NGC 4579 for which the point source associated with the nucleus of the
galaxy was excluded from the luminosity function. Also NGC 4945 is a
case where there is contribution to the X-ray emission from an
AGN. However the AGN is heavily obscured and the emission below about
10 keV of the AGN negligible \citep{schurch:02}.

\subsection{Contribution of LMXBs}
\label{sec:lmxb}

Due to the absence of optical identifications of a donor star in the
X-ray binaries detected by CHANDRA in other galaxies, except for LMC
and SMC, there is no obvious way to discriminate the contribution of
low mass X-ray binaries. On the other hand the relative contribution
of LMXB sources can be estimated and, as it was mentioned above, it
was one of the requirements to minimise the LMXB contribution, that
determined our selection of the late-type/starburst galaxies.

Due to the long evolution time-scale of LMXBs we expect the population
of LMXB sources to be roughly proportional to the stellar mass of a
galaxy, whereas the population of short-living HMXBs should  be
defined by the very recent value of the star formation rate. Therefore
the relative importance of LMXB sources should be roughly
characterised by (inversely proportional to) the ratio of star
formation rate to stellar mass of a galaxy. Since the determination of
stellar mass, especially for a starburst galaxy, is very difficult and
uncertain we used values for the total mass of a galaxy estimated from
dynamical methods and assumed that the stellar mass is roughly
proportional to the total mass.
To check our assumption we compare the dynamical mass with the
$K$ band luminosity for galaxies for which, first, enough data exist to
construct a growth curve in the $K$ band and, second, for which an
extrapolation to the total K band flux can be made following the
approach of \citet{spinoglio:95}. The number of galaxies is small, the
sample consists of M 74, M 83, NGC 4736 and NGC 891, and the
uncertainties associated with this approach are big, i.e. of order a
factor 3. But within this uncertainty there is a correlation between
the $K$ band luminosity and the dynamical mass estimate. However, due to
the more abundant data for and higher accuracy of dynamical masses we
do not use stellar mass estimates based on $K$ band luminosities in the
following.
The values of the total dynamical mass,
corresponding references, and the ratios of SFR to total mass are
given in Table \ref{tab:gal1} and Table \ref{tab:gal2}.

The SFR to total mass ratios for late-type galaxies should be compared
with that for the Milky Way, for which the population of sufficiently
luminous X-ray binaries is studied rather well \citep{grimm:02}. We
know that the Milky Way, having a ratio SFR/M$_{dyn}$ $\sim
5\cdot10^{-13}$ yr$^{-1}$, or SFR/M$_{stellar}$ $\sim 5\cdot10^{-12}$
yr$^{-1}$, is dominated by LMXB sources, HMXB sources contributing
$\sim$ 10 per cent to the total X-ray luminosity and $\sim$ 15 per
cent to the total
number (above $\sim 10^{37}$ erg s$^{-1}$) of X-ray binaries. As can be seen
from Table \ref{tab:gal1}, concerning the galaxies for which
luminosity functions are available the minimal value of SFR/M$_{dyn}$
$\sim 1.5\cdot 10^{-11}$ yr$^{-1}$ is achieved for M 74 and NGC 4736,
which exceeds by a factor of $\sim 30$ that of the Milky
Way. Therefore, even in the least favourable case of these two
galaxies, we expect the HMXB sources to exceed LMXBs by a factor of
$\sim 3$ at least, both in number and in luminosity. A more detailed
comparison is shown in Fig. \ref{fig:lmxb}, where we plot the expected
contributions of LMXBs and HMXBs to the observed luminosity function
for NGC 4736. The luminosity function of HMXBs was obtained by scaling
the Milky Way HMXB luminosity function by the ratio of SFRs of NGC
4736 to the Milky Way. The LMXB contribution was similarly estimated
by scaling the Milky Way LMXB luminosity function by the ratio of the
corresponding total masses. As can be seen from Fig. \ref{fig:lmxb},
the contribution of LMXB sources does not exceed $\sim$ 30 per cent at the
lower luminosity end of the luminosity function. If the fractions of
NSs and BHs in low mass systems in
late-type/starburst galaxies are similar to that in the Milky Way, the
contribution of LMXBs should be negligible at luminosities above $\sim
10^{38}$ erg s$^{-1}$, corresponding to the Eddington limit of a neutron
star, to which range most of the following analysis will be
restricted.

For all galaxies from Tables \ref{tab:gal1} and \ref{tab:gal2} the
lowest values for SFR/M are $4\cdot 10^{-12}$ and $9\cdot 10^{-12}$ for
IC 342 and NGC 891, respectively. This means that the contribution of
LMXBs could make up a sizeable portion of their X-ray luminosity,
$\sim$50 per cent for IC 342 and $\sim$25 per cent for NGC 891.
Therefore their data points should be considered as upper limits
on the integrated luminosity of HMXBs (shown in Fig.  \ref{fig:l-sfr}
as arrows).

\begin{figure}
  \resizebox{\hsize}{!}{\includegraphics{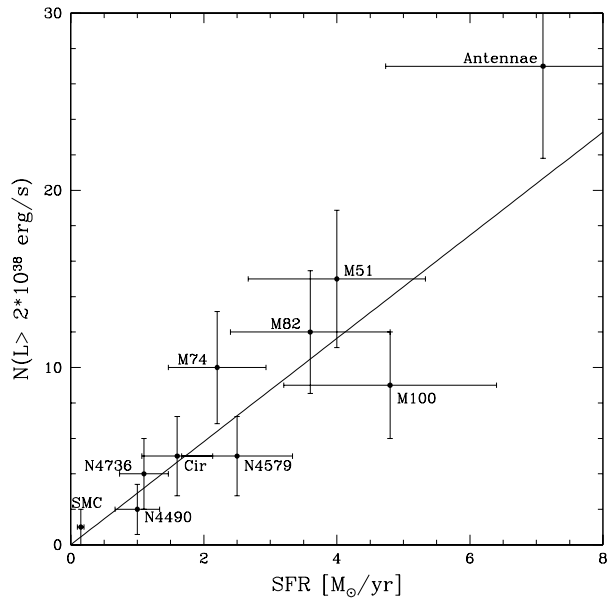}}
  \caption{Number of sources with a 2--10 keV luminosity exceeding $2
  \cdot 10^{38}$ erg s$^{-1}$ versus SFR for galaxies from Table
  \ref{tab:gal1}. The figure shows a clear correlation between the
  number of sources and the SFR. The straight line is the best fit
  from a Maximum-Likelihood fit, Eq. \ref{eq:number}. The vertical
  error bars were calculated assuming a Poissonian distribution, the
  SFR uncertainty was assumed to be 30 per cent. For M 74 and M 100, whose
  completeness limit exceeds $2\cdot10^{38}$ erg s$^{-1}$ the contribution of
  sources above $2\cdot10^{38}$ erg s$^{-1}$ and below the completeness limit
  was estimated from the ``universal'' luminosity function,
  Eq. \ref{eq:bestc}.}
\label{fig:numsfr}
\end{figure}

\section{High Mass X-ray Binaries as a star formation indicator}

As already mentioned, the simplest assumption about the connection of
HMXBs and SFR would be that the number of X-ray sources with a high
mass companion is directly proportional to the star formation rate in
a galaxy. In Fig. \ref{fig:lfs} (right panel) we show the luminosity
functions of the galaxies from our primary sample scaled to the star
formation rate of the  Antennae galaxies. Each luminosity function is
plotted above its corresponding completeness limit. It is obvious that
after rescaling the luminosity functions occupy a rather narrow band
in the log(N)-log(L) plane and seem to be consistent with each other
within a factor of $\sim2$ whereas the star formation rates differ by
a factor of $\sim40-50$. This similarity indicates that the
number/luminosity function of HMXB sources might indeed be
proportional to the star formation rate. This conclusion is further
supported by Fig. \ref{fig:numsfr} which shows the number of sources
with a luminosity above $2 \cdot 10^{38}$ erg s$^{-1}$ versus the SFR. The
threshold luminosity was chosen at  $2 \cdot 10^{38}$ erg s$^{-1}$ to have a
sufficient number of galaxies with a completeness limit equal or lower
than that value and, on the other hand, to have a sufficient number of
sources for each individual galaxy. In addition, as was discussed in
Sec. \ref{sec:lmxb}, this choice of the threshold luminosity might
help to minimise the contribution of LMXB sources. The errors for the
number of sources were computed assuming a Poissonian distribution.
For the SFR values we assumed a 30 per cent uncertainty. Although the errors
are rather big, the correlation of the number of sources with SFR is
obvious. The slope of the correlation, determined from a least-squares
fit in the form $N \propto SFR^{\alpha}$, is $\alpha= 1.06 \pm 0.07$,
i.e. it is consistent with unity. A fit of this correlation with a
straight line  $N \propto SFR$ (shown in the figure by solid line)
gives:
\begin{eqnarray}
N(L>2 \cdot 10^{38}\text{erg s$^{-1}$}) = (2.9 \pm 0.23) \cdot SFR
\text{[M$_{\odot}$ yr$^{-1}$]}
\label{eq:number}
\end{eqnarray}
According to this fit we should expect less than 1 source in the
Milky Way, having a SFR of 0.25 M$_{\odot}$ yr$^{-1}$, which is in agreement
with the fact that no source above this luminosity is observed
\citep{grimm:02}.

\subsection{Universal HMXB Luminosity Function ?}
\label{sec:ulf}

\begin{figure}
  \resizebox{\hsize}{!}{\includegraphics{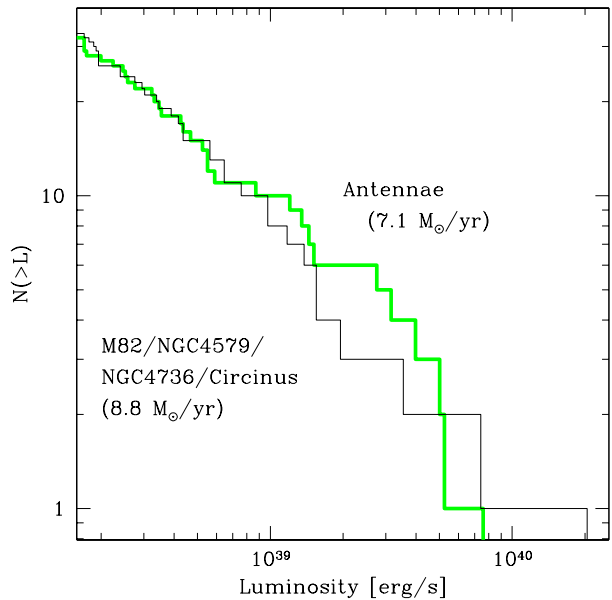}}
  \caption{Comparison of the combined luminosity function of M 82, NGC
  4579, NGC 4736 and Circinus, having SFRs in the range 1--3.5
  M$_{\odot}$ yr$^{-1}$ with the Antennae luminosity function (7.1
  M$_{\odot}$ yr$^{-1}$). A Kolmogorov-Smirnov test gives a probability of
  15 per cent that the two luminosity functions are derived from the same
  distribution. See discussion in the text regarding the effect of the
  errors in the distance measurements on the shape of the combined
  luminosity function.}
\label{fig:comp_l}
\end{figure}

In order to check the assumption that all the individual luminosity
functions have identical or similar shape with the normalisation being
proportional to the SFR, we compare the luminosity function of the
Antennae galaxies, having a high star formation rate ($\sim$ 7
M$_{\odot}$ yr$^{-1}$), with the collective luminosity function of galaxies
with medium SFRs (in the range of $\sim$ 1.0-3.5 M$_{\odot}$ yr$^{-1}$). For
the later we summed the luminosity functions of M 82, NGC 4579, NGC
4736 and Circinus, having a combined SFR of $\sim$ 8.8
M$_{\odot}$ yr$^{-1}$. The two luminosity functions (shown in
Fig. \ref{fig:comp_l}) agree very well at $L_X \la 10^{39}$ erg s$^{-1}$ with
possible differences at higher luminosities. In a strict statistical
sense, a Kolmogorov-Smirnov test gives a 15 per cent probability that the
luminosity functions are derived from the same distribution, thus,
neither confirming convincingly, nor rejecting the null hypothesis.
However, it should be emphasised, that whereas the shape of a single
slope power law luminosity function is not affected at all by the
uncertainty in the distance, more complicated forms of a luminosity
function, e.g. a power law with cut-off, would be sensitive to errors
in the distance determination. The effect might be even stronger for
the combined luminosity functions of several galaxies, located at
different distances and each having different errors in the distance
estimate. In the case of a power law with high luminosity cut-off, the
effect would be strongest at the high luminosity end and will
effectively dilute the cut-off, as probably is observed. Therefore, we
can presently not draw a definitive conclusion about the existence of
a universal luminosity function of HMXBs, from which all luminosity
functions of the individual galaxies are {\it strictly} derived. For
instance, subtle effects similar to the effect of flattening of the
luminosity function with increase of SFR suggested by
\citet{kilgard:01,ghosh:01,ptak:01} can not be excluded based on the
presently available sample of galaxies and sensitivities achieved. We
can conclude, however, that there is no evidence for strong non-linear
dependences of the luminosity function on the SFR.

\begin{figure*}
  \resizebox{0.49\hsize}{!}{\includegraphics{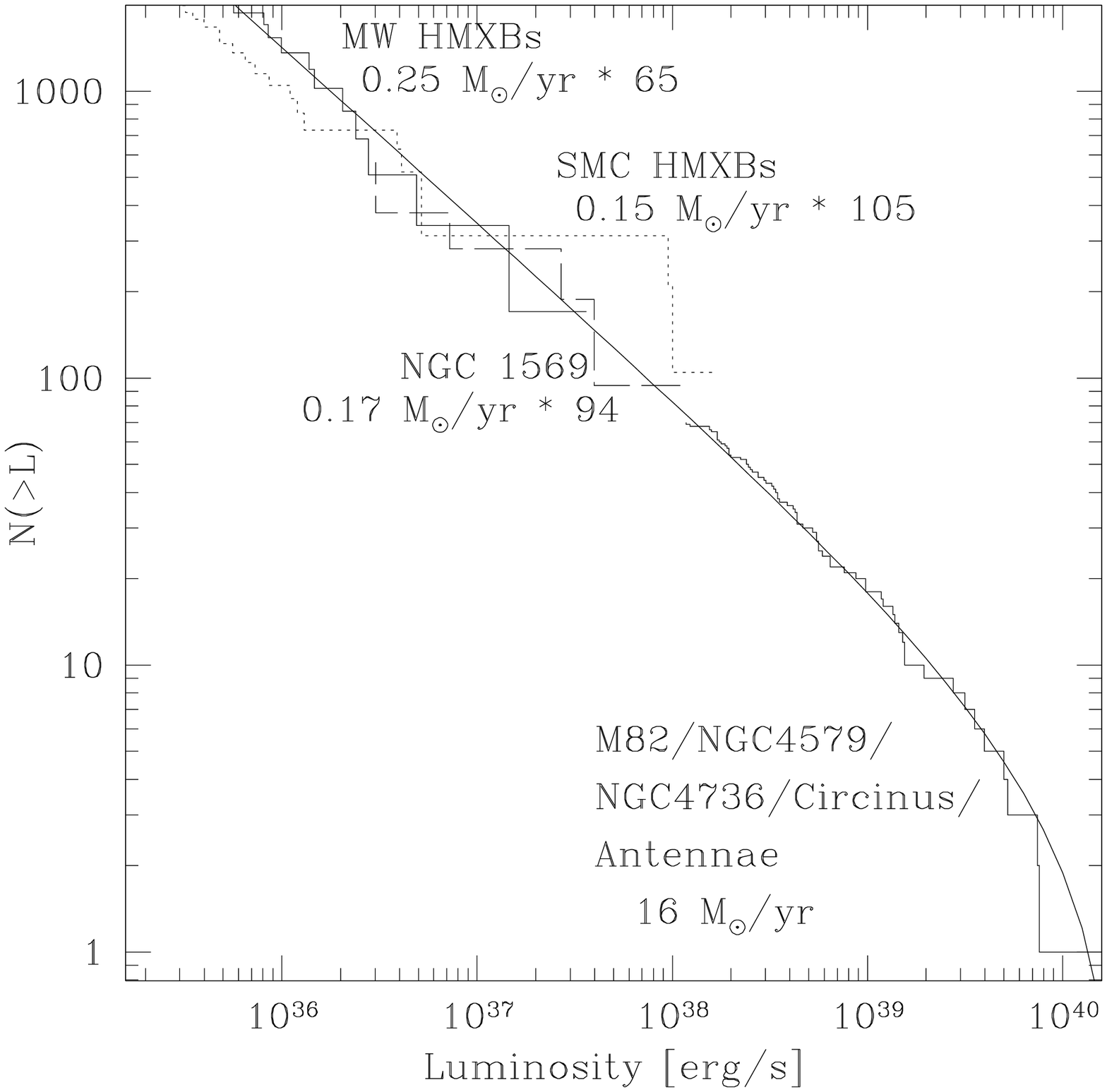}}
  \resizebox{0.49\hsize}{!}{\includegraphics{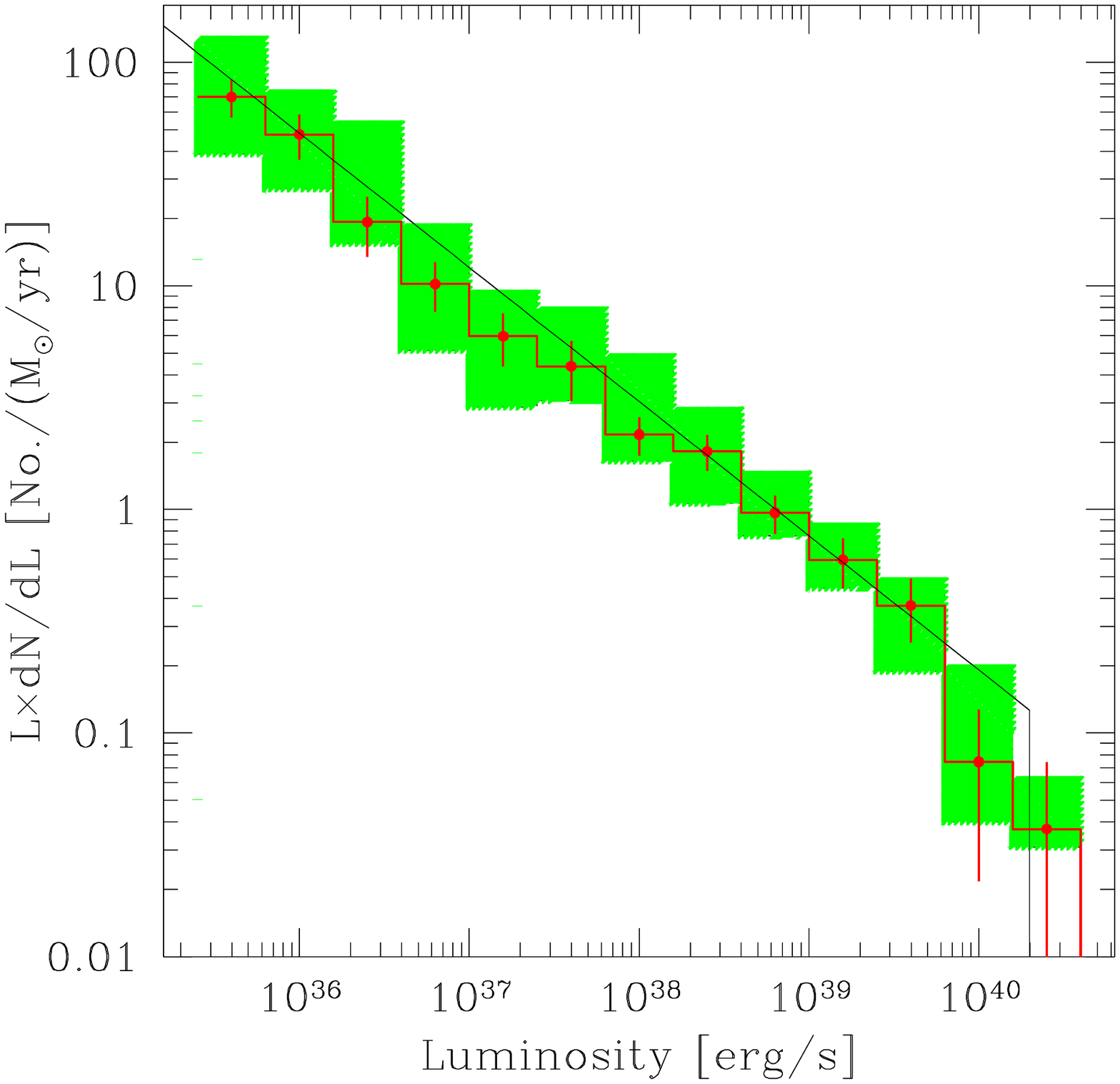}}
  \caption{{\em Left:} Combined luminosity function of compact X-ray
  sources in the starburst galaxies M82, NGC 4038/9, NGC 4579, NGC
  4736 and Circinus with a total SFR of 16 M$_{\odot}$ yr$^{-1}$ (histogram
  above $2\cdot 10^{38}$ erg s$^{-1}$) and the luminosity functions of NGC
  1569 and HMXBs in the Milky Way and Small Magellanic Clouds (three
  histograms below $2\cdot 10^{38}$ erg s$^{-1}$). The thin solid line is the
  best fit to the combined luminosity function of the starburst
  galaxies {\em only}, given by Eq. \ref{eq:bestc}. {\em Right:}
  Differential luminosity function obtained by combining the data for
  {\em all} galaxies from the primary sample, except for NGC 3256 (see
  text). The straight line is the best fit to the luminosity function
  of star forming galaxies given by Eq. \ref{eq:bestd}. -- the same as
  in the left hand panel. Note, that due to different construction
  algorithms, the luminosity functions shown in the left and right
  panels are based on different but overlapping samples of galaxies
  (see discussion in the text). The grey area is the 90 per cent confidence
  level interval we obtained from a Monte-Carlo simulation taking into
  account uncertainties in the SFR and distances. For details see
  discussion in the text.}
\label{fig:all_l}
\end{figure*}

\begin{figure*}
  \resizebox{\hsize}{0.9\vsize}{\includegraphics{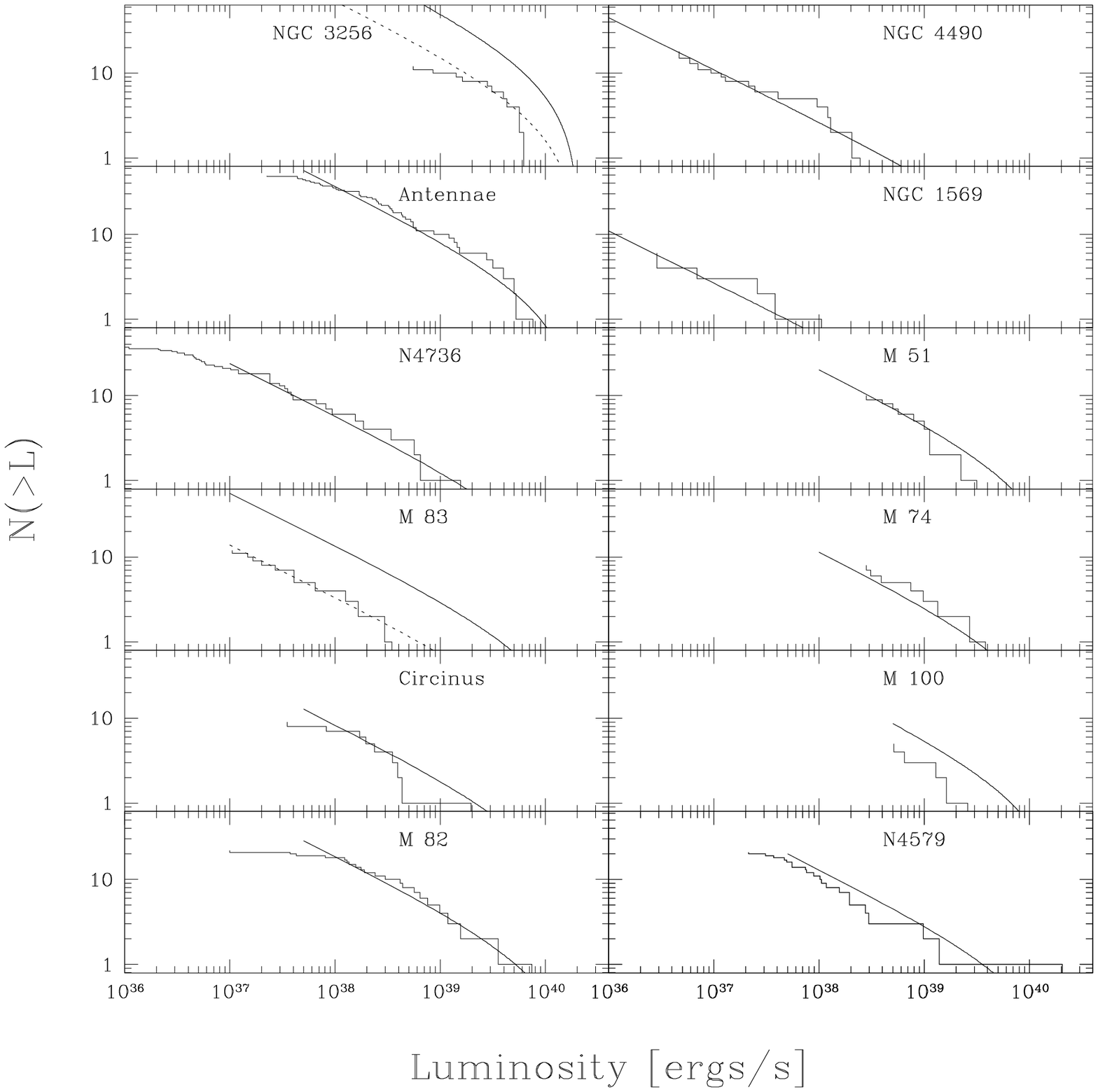}}
  \caption{Comparison of the ``universal'' luminosity function defined
  by Eq. \ref{eq:bestd} (thin solid lines) with individual luminosity
  functions of compact X-ray sources in the galaxies from Table
  \ref{tab:gal1} (histograms). The normalisation of the ``universal''
  luminosity function in each panel was calculated using corresponding
  SFR values from Table \ref{tab:gal1}. For M83 the luminosity
  function of the compact sources in the nuclear region only is
  plotted, whereas the normalisation of the ``universal'' luminosity
  function was computed using the overall SFR for the
  galaxy. Therefore the thin line should be considered as an upper
  limit. The dotted lines are fits to the normalisation of the
  observed luminosity functions in the cases where completeness or
  coverage do not represent the same area as the SFR measurements.}
\label{fig:all_lf}
\end{figure*}

As the next step we compare the luminosity functions of actively star
forming galaxies with that of low SFR galaxies. Unfortunately, the
X-ray binary population of low SFR galaxies is usually dominated
by LMXB systems. One of the cases in which the luminosity function of
HMXB sources can be reliably obtained is the Milky Way galaxy, for
which all sufficiently bright X-ray binaries are optically
identified. Another case is the Small Magellanic Cloud, which has a
SFR value similar to our Galaxy, but is $\sim 300-500$ less massive
and, correspondingly, has very few, if any, LMXB sources
\citep{yokogawa:00}. Moreover, the SMC is close enough to have
optical identifications of HMXBs which makes a distinction like in the
Milky Way possible. In order to do the comparison, we combined the
luminosity functions of all actively star forming galaxies from our
sample with a completeness limit lower than $2\cdot 10^{38}$ erg s$^{-1}$
-- M 82, Antennae,  NGC 4579, NGC 4736 and Circinus. These galaxies
have a total SFR of $\sim$ 16 M$_{\odot}$ yr$^{-1}$, which exceeds the Milky
Way SFR ($\sim 0.25$  M$_{\odot}$ yr$^{-1}$) by a factor of $\sim 65$.
Fig. \ref{fig:all_l} shows the combined luminosity function of the 
above mentioned star forming galaxies and the luminosity functions of
Galactic and SMC HMXBs scaled according to the ratios of SFRs.
Shown in Fig. \ref{fig:all_l} by a solid line is the fit to the
luminosity function of the high SFR galaxies {\em only} (see below),
extrapolated to lower luminosities. It is obvious that the luminosity
functions of Galactic and SMC HMXBs agree surprisingly well with an
extrapolation of the combined luminosity function of the starburst
galaxies.

Thus we demonstrated that the presently available data are consistent
with the assumption that the {\it approximate} shape and normalisation
of the luminosity function for HMXBs in a galaxy with a known star
formation rate can be derived from a ``universal'' luminosity function
whose shape is fixed and whose normalisation is proportional to
star formation rate. Due to a number of uncertainties involved, the
accuracy of this approximation is difficult to assess. Based on our
sample of galaxies we can conclude that it might be accurate within
$\sim$ 50 per cent.

In order to obtain the universal luminosity function of HMXBs we fit
the combined luminosity function of M 82, Antennae, NGC 4579, NGC 4736
and Circinus using a Maximum-Likelihood method with a power law with
a cut-off at $L_c = 2.1\cdot10^{40}$ erg s$^{-1}$ and normalise the result to
the combined SFR of the galaxies. The best fit luminosity function
(solid line in Fig.\ref{fig:all_l}) in the differential form is given
by:
\begin{equation}
\frac{dN}{dL_{38}}= (3.3^{+1.1}_{-0.8})\cdot
                  SFR \cdot L_{38}^{-1.61\pm0.12}
  \text{~~for } L<L_c,
\label{eq:bestd}
\end{equation}
where $L_{38}=L/10^{38}$ erg s$^{-1}$ and SFR is measured in units of 
M$_{\odot}$ yr$^{-1}$. The errors are $1 \sigma$ estimates
for one parameter of interest. The rather large errors for
normalisation are due to the correlation between slope and
normalisation of the luminosity function, with a higher value of 
normalisation corresponding to a steeper slope. The cumulative form of
the luminosity function, corresponding to the best values of the slope
and normalisation is:
\begin{equation}
N(>L)= 5.4 \cdot SFR \cdot
         (L_{38}^{-0.61} -  210^{-0.61}),
\label{eq:bestc}
\end{equation}
According to a Kolmogorov-Smirnov test the data are consistent with
the best fit model at a confidence level of 90 per cent.

As an additional test we checked all individual luminosity functions
against our best fit using a Kolmogorov-Smirnov test. Taking into
account the respective completeness limits, the shapes of all
individual luminosity functions are compatible with the assumption of
a common 'origin'. In Fig. \ref{fig:all_lf} we show the individual
luminosity functions along with the universal luminosity function
given by Eq.(\ref{eq:bestd}) with the normalisation determined
according to the corresponding star formation rates derived from the
conventional SFR indicators (Table \ref{tab:gal1}).

Finally, we construct the differential luminosity function combining
the data for all galaxies from the primary sample, except for
NGC 3256 (having somewhat uncertain completeness limit). To do so we
bin all the sources above the corresponding completeness limits in
logarithmically spaced bins and normalise the result by the combined
SFR of all galaxies contributing to a given bin. Such a method
has the advantage of using all the available data. A disadvantage is
that due to significantly different luminosity ranges of the
individual luminosity functions (especially SMC and Milky Way on one
side and star forming galaxies on the other) uncertainties in the
conventional SFR estimates may lead to the appearance of artificial
features in the combined luminosity function. With that in mind, we
plot the differential luminosity function in the right panel of
Fig. \ref{fig:all_l} along with the best fit power law from
Eq.(\ref{eq:bestd}).

In order to investigate the influence of systematic uncertainties
in SFR and distance we performed a Monte-Carlo simulation taking into
account these two effects. The grey area in the right panel of
Fig. \ref{fig:all_l} shows the 90 per cent confidence interval obtained
from this simulation. In the simulation we randomly varied the
distances of galaxies, assuming the errors on the distance to be
distributed according to a Gaussian with a mean of 0 and a width of
20 per cent of the distance of a galaxy which corresponds to an uncertainty
in luminosity of $\sim$ 40 per cent. Correspondingly the SFR, affected in the
same way as the X-ray luminosity by uncertainties in the distance, was
changed. Additionally the SFR was randomly varied
also assuming a Gaussian error distribution with a mean of 0 and a
width of 30 per cent of the SFR, as assumed for Fig. \ref{fig:numsfr}. For
the Milky Way we varied in each Monte-Carlo run the distance to each
HMXB independently with a Gaussian with a mean of 0 and a width of
20 per cent of the distance.

Noteworthy is the fact that the luminosity function is sufficiently
close to a single slope power law in a broad luminosity range covering
more than five orders of magnitude. If the absence of significant
features is confirmed this allows to constrain the relative abundance
of NS and BH binaries and/or the properties of accreting compact
objects at supercritical accretion rates (see discussion in
Sec. \ref{sec:nsbh}).

However, it should be emphasised that there is hardly any overlap
in the luminosity functions for low and high SFR galaxies, as is
obvious from Fig. \ref{fig:lfs} and \ref{fig:all_l}. It happens that
this gap is around the 
Eddington luminosity of a NS, $L_{Edd, NS}$, which should be a dividing
line between NS and BH binaries. From simple assumptions it would be
expected that the luminosity functions below $L_{Edd, NS}$ are
dominated by NS whereas above $L_{Edd, NS}$ BH binaries should
dominate. This would imply a break in the luminosity function around
$L_{Edd, NS}$ because of different abundances of NSs and BHs. Due to
the uncertainties in SFR measurements it is possible that a break,
that would theoretically be expected around $L_{Edd, NS}$, could be
hidden by this gap. Even upper limits (not more than twice) are of
importance and could give some additional information about the
relative strength of the two populations of accreting binaries (see
discussion in Sec. \ref{sec:disc}). Observations of star forming
galaxies with sufficient sensitivity, i.e. with a completeness limit
well below $10^{38}$ erg s$^{-1}$ will be able to resolve this question.

\subsection{High Luminosity cut-off}
\label{sec:cutoff}

The combined luminosity function shown in the left panel of
Fig. \ref{fig:all_l} indicates a possible presence of a cut-off at
$L_c\sim 2\cdot10^{40}$ erg s$^{-1}$. From a statistical point of view,
when analysing the combined luminosity function of the high SFR
galaxies only, the significance of the cut-off is not very high, with
a single slope power law with slope 0.74 for the cumulative luminosity
function also giving an acceptable fit, although with a somewhat lower
probability of 54 per cent according to a Kolmogorov--Smirnov test.
However, an independent strong evidence for the existence of a cut-off
around $\sim {\rm few} \cdot 10^{40}$ erg s$^{-1}$ is provided by the
$L_X$--SFR relation as discussed in the next subsections.

The existence of such a cut-off, if it is real and if it is a
universal feature of the HMXB luminosity function, can have
significant implications to our understanding of the so-called
ultra-luminous X-ray sources. Assuming that these very luminous objects
are intermediate mass BHs accreting at the Eddington limit,
the value of the cut-off gives an upper limit on the mass of the black
hole, $\sim 100$ M$_{\odot}$. These apparently super-Eddington
luminosities can also be the result of other effects, like  a strong
magnetic field in NSs which may allow radiation to escape
without interacting with the accreting material \citep{basko:76},
emission from a supercritical accretion disk
\citep{shakura:73,paczynski:80}, beamed emission \citep{king:01}, or
the emission of a jet as suggested by \citet{koerding:02}. Moreover,
in BHs in high state radiation is coming from the quasi-flat
accretion disk where electron scattering gives the main contribution
to the opacity. It is easy to show that the radiation flux
perpendicular to the plane of the disk exceeds the average value by up
to 3 times \citep{shakura:73}. Also the Eddington luminosity is
dependent on chemical abundance which allows a twice higher luminosity
for accretion of helium. These last two effects alone can provide a
factor of 6 above the canonical Eddington luminosity.

It should be mentioned that, based on the combined luminosity function
only, we can not exclude the possibility that the cut-off is 
primarily due to the Antennae galaxies which contributes about half
of the sources above $10^{39}$ erg s$^{-1}$ and shows a prominent cut-off
in its luminosity function.  On the other hand, further indication for
a cut-off is provided by the luminosity function of NGC 3256.
Conventional star formation indicators give a value of SFR of $\sim$
45 M$_{\odot}$ yr$^{-1}$, however its luminosity function also shows a
cut-off at $\sim 10^{40}$ erg s$^{-1}$. Unfortunately, due to the large
distance (35 Mpc) and a comparatively short exposure time of the
CHANDRA observation, $\sim$ 28 ks, the luminosity function of NGC 3256
becomes incomplete at luminosities shortly below the brightest source
and therefore does not allow for a detailed investigation.

\subsection{Total X-ray luminosity as SFR indicator}
\label{sec:total}

CHANDRA and future X-ray missions with angular resolution of the order
of $\sim 1''$ would be able to spatially resolve X-ray binaries only
in nearby galaxies ($d \la \sim 50-100$ Mpc). For more distant
galaxies only the total luminosity of a galaxy due to HMXBs can be
used for X-ray diagnostics of star formation.

\begin{figure}
  \resizebox{\hsize}{!}{\includegraphics{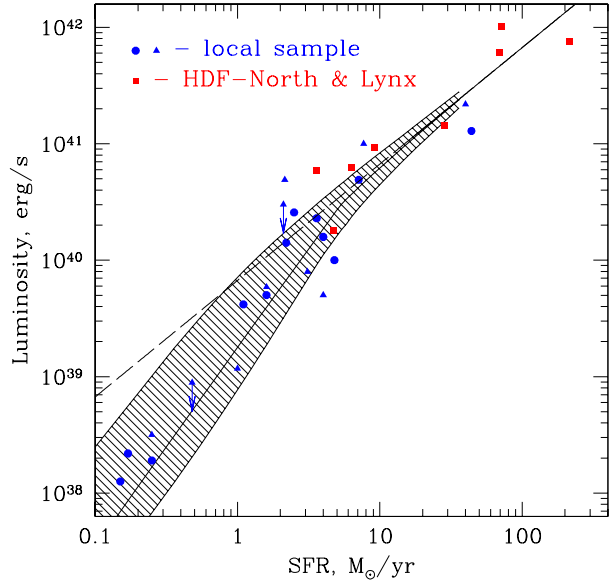}}
  \caption{The $L_X$--SFR relation. The filled circles and triangles
  are nearby galaxies from Table \ref{tab:gal1} (primary sample) and
  Table \ref{tab:gal2} (secondary sample), the open circles are
  distant star forming galaxies from the HDF North and Lynx field.
  The arrows are the upper limits for the X-ray luminosity due to
  HMXBs for IC 342 and NGC 891. The thick solid line shows the
  expected relation between SFR and the most probable value of the
  total luminosity computed for the best fit parameters of the HMXB
  luminosity function (exact calculation, from \citet{gilfanov:02}).
  Note, that in the low SFR regime the probability to find a galaxy
  below the solid curve is $\sim 10-15$ per cent. The shaded area shows the
  68 per cent confidence region including both intrinsic variance of the
  $L_X$--SFR relation and uncertainty of the best fit parameters of
  the HMXB luminosity function (Eq.(\ref{eq:bestd})). The dashed line
  shows the linear $L_X$--SFR relation given by Eq.(\ref{eq:ltot}).}
\label{fig:l-sfr}
\end{figure}

Fig. \ref{fig:l-sfr} shows the total luminosity of X-ray binaries
(above $10^{36}$ erg s$^{-1}$) plotted versus SFR.  The galaxies from the
primary sample (listed in Table \ref{tab:gal1}) are shown by filled
circles. The galaxies, for which only total luminosity is available
(Table \ref{tab:gal2}) are shown as filled triangles. The luminosities
of the galaxies from the primary sample were calculated by summing the
luminosities of individual sources down to the completeness limit of
the corresponding luminosity function. The contribution of the sources
below the completeness limit was approximately accounted for by
integrating a power law distribution with slope $1.6$ and
normalisation obtained from the fit to the observed luminosity
function. Note, that due to the shallow slope of the luminosity
function the total luminosity depends only weakly on the lower
integration limit. 

As an additional data point we take luminosity and SFR for the Large
Magellanic Cloud. The SFR is similar to the Milky Way SFR
\citep{holtzman:99}. Since no luminosity function is presently
available for LMC we estimated its integrated X-ray luminosity as a
sum of the time averaged luminosities of the three brightest HMXB
sources (LMC X-1, X-3, X-4) as measured by ASM \citep{grimm:02},
$L_{2-10~\text{keV}}\approx 3.4\cdot 10^{38}$ erg s$^{-1}$. Contribution of
the weaker sources should not change this estimate significantly,
since the luminosity of the next brightest source is by a factor of
$\sim 30-50$ smaller \citep{sunyaev:90}.

\subsection{Theoretical $\bf L_X$--SFR relation}
\label{sec:lx-sfr}

At first glance, the relation between collective luminosity of HMXBs
and SFR can be easily derived integrating Eq. (\ref{eq:bestd}) for the
SFR dependent luminosity function. Therefore, as the population of
HMXB sources in a galaxy is directly proportional to SFR, one might
expect that the X-ray luminosity of galaxies due to HMXB, $L_X$,
should be linearly proportional to SFR. However this problem contains
some subtleties related to the statistical properties of the power law
luminosity distribution of discrete sources which appear not to have
been recognised previously (at least in astrophysical context).
The difference between the most probable value of the total luminosity
of HMXB sources in a galaxy (the mode of the distribution) and the
ensemble average value (expectation mean, obtained by integrating
Eq. (\ref{eq:bestd})) results in the non-linear $L_X$--SFR dependence
in the low SFR regime. As this effect might be of broader general
interest and might work in many different situations related to
computing/measuring integrated luminosity of a limited number of
discrete objects, we give it a more detailed and rigorous discussion
in a separate paper \citep{gilfanov:02}, and restrict the discussion
here to only a brief explanation and an approximate analytical
treatment. A somewhat similar problem was considered by
\citet{kalogera:01} in the context of pulsar counts and the faint end
of the pulsar luminosity function.

For illustration only, let us consider a population of discrete
sources with a Gaussian luminosity function. As is well known, in this
case the sum of their luminosities -- the integrated luminosity of the
parent galaxy, also obeys a Gaussian distribution for which the mean
luminosity and dispersion can be computed straightforwardly. An
essential property of this simple case is that for an ensemble of
galaxies, each having a population of such sources, the most probable
value of the integrated luminosity of an arbitrarily chosen galaxy 
(the mode of the distribution) equals to the mean luminosity (averaged
over the ensemble of galaxies). The situation might be different in
the case of a population of discrete sources with a power law (or
similarly skewed) luminosity function. In this case an ensemble of
galaxies would have a non-Gaussian probability distribution of the
integrated luminosity. Due to skewness of the probability distribution
in this case, the most probable value of the integrated luminosity of
an arbitrarily chosen galaxy does not necessarily coincide with the
mean value (the ensemble average). The effect is caused by the fact
that depending on the slope of the luminosity function and its
normalisation the integrated luminosity of the galaxy might be defined
by a small number of brightest sources even when the total number of
sources is large. Of course, in the limit of large number of sources
in the high luminosity end of the luminosity function the distribution
becomes asymptotically close to Gaussian and, correspondingly, the
difference between the most probable value and the ensemble average
vanishes. In this limit the relation between the integrated luminosity
of HMXBs and SFR can be derived straightforwardly integrating
Eq.(\ref{eq:bestd}) for $L_c = 2.1\cdot10^{40}$ erg s$^{-1}$:
\begin{equation}
L_X=6.7\cdot 10^{39}\cdot SFR \text{[M$_{\odot}$ yr$^{-1}$]}
\text{~~erg s$^{-1}$}
\label{eq:ltot}
\end{equation}
It should be emphasised that the ensemble average integrated luminosity
(i.e. averaged over many galaxies with similar SFR) is always
described by the above equation, independent of the number of
sources and shape of the luminosity function. This equality is
maintained due to the outlier galaxies, whose luminosity exceeds
significantly both the most probable and average values. These outlier
galaxies will result in enhanced and asymmetric dispersion in the
low SFR-regime.

The following simple consideration leads to an approximate analytical
expression for the most probable value of the integrated luminosity.
Assuming a power law luminosity function $dN/dL=A\cdot SFR \cdot
~L^{-\alpha}$ with $1<\alpha<2$, one might expect,  that the brightest
source would most likely have a luminosity $L_{max}$ close to the
value $\sim L_1$ such that $N(>L_1)\sim 1$, i.e.
\begin{eqnarray}
L_1\propto SFR^{\frac{1}{\alpha-1}}
\end{eqnarray}
In the presence of a cut-off $L_{c}$ in the luminosity function, the
luminosity of the brightest source, of course, can not  exceed the
cut-off luminosity: $L_{max}=\text{min}(L_1,L_{c})$. The most probable
value of the total luminosity can be computed integrating the
luminosity function from $L_{min}$ to $L_{max}=\text{min}(L_1,L_{c})$:
\begin{eqnarray}
  L_{total} = \int^{\text{min}(L_1,L_{c})}_{L_{min}}
  \frac{dN}{dL} L~dL,
\end{eqnarray}
which leads to
\begin{eqnarray}
\label{eq:ltot_sfr}
  L_{total} \approx \frac{A\cdot SFR}{2 - \alpha} \cdot
  \text{min}(L_1,L_{c})^{2 - \alpha}
\end{eqnarray}
for $1 < \alpha < 2$ and $L_1,L_{c} >> L_{min}$.

Obviously there are two limiting cases of the $L_X$--SFR dependence of
the total luminosity on SFR, depending on the relation between $L_c$
and $L_1$, i.e. on the expected number of sources in the high end of
the luminosity function, near its cut-off. In the limit of low SFR
(small number of sources) $L_1<L_c$ and the luminosity of the
brightest source would increase with SFR: $L_{max}\sim L_1\propto
SFR^{\frac{1}{\alpha-1}}$. Therefore the $L_X$--SFR dependence might
be strongly non-linear:
\begin{eqnarray}
  L_{total} \propto SFR^{\frac{1}{\alpha-1}}
\label{eq:low_sfr}
\end{eqnarray}
e.g. for $\alpha=1.5$ the relation is quadratic $L_{total} \propto
SFR^2$. For sufficiently large values of SFR $L_1>L_c$,
i.e. $N(>L_c)>1$ implying a large number of sources in the high
luminosity end of the luminosity function and, correspondingly,
Gaussian probability distribution of the integrated luminosity.
In this case $L_{max}\sim L_c=const$ and does not depend on SFR
anymore and the dependence is linear, in accord with
Eq.(\ref{eq:ltot}).

Importantly, the entire existence of the linear regime in the
$L_X$--SFR relation is a direct consequence of the existence of a
cut-off in the luminosity function. For a sufficiently flat luminosity
function,  $1 < \alpha < 2$, the collective luminosity of the sources
grows faster than linear because brighter and brighter sources
define the total luminosity as the star formation rate increases. Only
in the presence of the maximum possible luminosity of the sources,
$L_c$ (for instance Eddington limit for NSs) the regime can
be reached, when $N(>L_c)$ becomes larger than unity and subsequent
increase of the star formation rate results in the linear growth of
the total luminosity. The latter, linear, regime of the $L_X$--SFR
relation was studied independently by \citet{ranalli:02} based on ASCA
and Beppo-SAX data. Note that their equation (12) agrees with our
Eq.(\ref{eq:ltot}) within 15 per cent.

The position of the break in the $L_X$--SFR relation depends on the
slope of the luminosity function and the value of the cut-off
luminosity:
\begin{eqnarray}
{\rm SFR}_{\rm break}\propto L_c^{\alpha-1}
\end{eqnarray}
Combined with the slope of the $L_X$--SFR relation in the low SFR
regime (Eq.(\ref{eq:low_sfr})) and the normalisation of the linear
dependence in the high SFR limit this opens a possibility to constrain
the parameters of the luminosity function studying the $L_X$--SFR
relation alone, without actually constructing luminosity functions,
e.g. in distant unresolved galaxies.

\begin{table*}
\caption{Sample galaxies from the Hubble Deep Field North and Lynx Field}
\begin{tabular}{|l|r|c|c|c|c|}
\hline
Source  & redshift & $F_{1.4~GHz}$ & SFR & $S_{0.5-8~keV}$ &$L_X$ \\
        &     &  [$\mu$Jy]    &[M$_{\odot}$ yr$^{-1}$] &[$10^{-15}$
erg s$^{-1}$ cm$^{-2}$]&[$10^{40}$ erg s$^{-1}$]\\
\hline
123634.5+621213 & 0.458 &  233  & 28 & 0.43 & 14.4\\
123634.5+621241 & 1.219 &  230  &213 & 0.3  & 75.9\\
123649.7+621313 & 0.475 &   49  & 8  & 0.15 & 2.5\\
123651.1+621030 & 0.410 &   95  & 9  & 0.3  & 9.3\\
123653.4+621139 & 1.275 &   66  &69  & 0.22 & 60.6\\
123708.3+621055 & 0.423 &   45  & 4  & 0.18 & 5.9\\
123716.3+621512 & 0.232 &  187  & 5  & 0.18 & 1.8\\
\hline
084857.7+445608 & 0.622 &  320  &71  & 1.46 &102 \\
\hline
\end{tabular}
\label{tab:hdfn}
\flushleft
For two galaxies, 123634.5+621213 and 123651.1+621030, there exist
stellar mass estimates obtained with the method of
\citet{brinchmann:00} of $4.2 \cdot 10^{11}$ M$_{\odot}$ and $7 \cdot
10^{10}$ M$_{\odot}$ respectively, which show that the galaxies are
dominated by HMXBs (J. Brinchmann, private communication).
\end{table*}

\subsection{$\bf L_X$--SFR relation: comparison with the data}

The solid line in Fig.\ref{fig:l-sfr} shows the result of the exact
calculation of the $L_X$--SFR relation from \citet{gilfanov:02}. The
relation was computed for the best fit parameters of the HMXB
luminosity function determined from the analysis of five mostly well
studied galaxies from the primary sample (section \ref{sec:ulf} and
Eq.(\ref{eq:bestd})). Note, that due to the skewness of the
probability distribution for $L_{total}$ in the non-linear, low SFR
regime the theoretical probability to find a galaxy below the most
probable value (the solid curve in  Fig.\ref{fig:l-sfr}) is $\approx
12-16$ per cent at SFR = 0.2-1.5 M$_{\odot}$ yr$^{-1}$ and increases
to $\approx$ 30 per cent at SFR = 4-5 ~M$_{\odot}$ yr$^{-1}$, near the
break of the $L_X$--SFR relation. In the linear regime (SFR $\ga 10$
M$_{\odot}$ yr$^{-1}$) it asymptotically approaches $\sim 50 per
cent$, as expected. The shaded area around the solid curve corresponds
to the 68 per cent confidence level including both intrinsic variance
of the  $L_X$--SFR relation and uncertainty of the best fit parameters
of the HMXB luminosity function (Eq.(\ref{eq:bestd})).

Fig.\ref{fig:l-sfr} demonstrates sufficiently good agreement between
the data and the theoretical $L_X$--SFR relation. Importantly, the
predicted relation agrees with the data both in the high and low SFR
regime, thus showing that the data, including the high redshift
galaxies from Hubble Deep Field North (see the following subsection),
are consistent with the HMXB luminosity function parameters, derived
from significantly fewer galaxies than plotted in Fig.\ref{fig:l-sfr}.

The existence of the linear part at SFR $>$ 5-10 M$_{\odot}$ yr$^{-1}$
gives an independent confirmation of the reality of the cut-off in the
luminosity function of HMXBs (cf. Sec. \ref{sec:cutoff}). The
position of the break and normalisation of the linear part in the
$L_X$--SFR relation confirms that the maximum luminosity of the HMXB
sources (cut-off in the HMXB luminosity function) is of the order of
$L_c\sim 10^{40}-10^{41}$ erg s$^{-1}$ (see \citet{gilfanov:02}
for more details). Despite the number of theoretical ideas being
discussed, the exact reason for the cut-off in the HMXB luminosity
function is not clear and significant variations of $L_c$ among
galaxies, related or not to the galactic parameters, such as
metalicity or star formation rate can not be excluded a priori.
However, significant variations in $L_c$ from galaxy to galaxy would
result in large dispersion in the break position and in the linear
part of the $L_X$--SFR relation. As such large dispersion is not
observed, one might conclude that there is no large variation of the
cut-off luminosity between galaxies and, in particular, there is
no strong dependence of the cut-off luminosity on SFR.

\subsection{Hubble Deep Field North}
\label{sec:hdfn}

In order to check whether the correlation, which is clearly seen from
Fig. \ref{fig:l-sfr} for nearby galaxies, holds for more distant
galaxies as well we used the data of the CHANDRA observation of the
Hubble Deep Field North \citep{brandt:01}. We cross-correlated the
list of the X-ray sources detected by CHANDRA with the catalogue of
radio sources detected by VLA at 1.4 GHz \citep{richards:00}. Using
optical identifications of \citet{richards:98} and redshifts from
\citet{cohen:00} we compiled a list of galaxies detected by CHANDRA
and classified as spiral or irregular/merger galaxies by
\citet{richards:98} and not known to show AGN activity. The
K-correction for radio luminosity was done assuming a power law
spectrum and using the radio spectral indices from
\citet{richards:00}. The X-ray luminosity was K-corrected and
transformed to the 2--10 keV energy range using photon indices from
\citet{brandt:01}. The final list of galaxies selected is given in
Table \ref{tab:hdfn}. An additional data point, X-ray flux and
redshift, is taken from the observation of the Lynx Field by
\citet{stern:02}. The radio flux is obtained from a cross-correlation
of the X-ray positions with \citet{oort:87}.

The star formation rates were calculated assuming that the non-thermal
synchrotron emission due to electrons accelerated in supernovae
dominates the observed 1.4 GHz luminosity and using the following
relation 
from \citet{condon:92}:
\begin{eqnarray}
\label{eq:ra2}
  &SFR_{radio} = 1.9 \cdot 10^{-29} \cdot
  \nu_{\text GHz}^{\alpha}\cdot L_{\nu} \text{[erg s$^{-1}$ Hz$^{-1}$]},
\end{eqnarray}
where $\alpha$ is the slope of the non-thermal radio emission.

The galaxies from HDF North and Lynx are shown in Fig.\ref{fig:l-sfr}
by open circles. A sufficiently good agreement with the
theoretical $L_X$--SFR relation is obvious.

\section{Discussion}
\label{sec:disc}
\subsection{Neutron stars, stellar mass black holes and intermediate
mass black holes} 
\label{sec:nsbh}

Two well known and one possible types of accreting objects should
contribute to the X-ray luminosity function of sources in star forming
galaxies:
\begin{enumerate}
\item neutron stars (M $\sim$ 1.4 M$_{\odot}$),
\item stellar mass black holes ($3\le M/M_{\odot}\le 20$) born due to
collapse of high mass stars, and
\item intermediate mass ($50 \la M/M_{\odot} \la 10^5$) black holes of
unknown origin.
\end{enumerate}
Each class of accreting objects is expected to have a maximum possible
luminosity, close or exceeding by a factor of several the
corresponding Eddington luminosity. In a general case we should expect
that each of these three types of accreting objects should have its
own luminosity function depending on the mass distribution inside each
class (more narrow for NSs, more broad for BHs and
probably very broad for intermediate mass BHs), properties of
the binary and mass loss type and rate from the normal star.
Therefore, the combined luminosity function of a galaxy, containing
all three types of objects should have several breaks or steps (see
Fig. \ref{fig:comp_lf}) which are not present in Fig. \ref{fig:all_l}.
Such breaks should be connected with the fact that, for example, below
the Eddington limit for a NS (or at somewhat higher
luminosity) more abundant NS X-ray binaries might dominate
in the number of objects, whereas at higher luminosities only black
holes should contribute due to their higher masses and broader mass
distribution. Until now CHANDRA data did not show any evidence for a
break in the luminosity function expected in the vicinity or above of
Eddington luminosity for NS mass. However, such a break must
exist, the only question is how pronounced and broad it is.

It is believed that stars with masses higher than 60-100 $M_{\odot}$
are unstable. Therefore there should be an upper limit on the mass
of BHs born as a result of stellar collapse. Until now the
most massive known stellar mass BH in our Galaxy, GRS
1915+105, has a mass of $\sim$15 M$_{\odot}$ \citep{greiner:01}. It is
natural that the Eddington luminosity of these objects, amplified
several times by angular distribution of radiation and chemical
abundance effects, should result in the maximum luminosity of X-ray
sources of this type. It is important to mention that 3 years of
RXTE/ASM observations revealed from time to time super-Eddington
luminosities of some Galactic X-ray binaries on the level of 3--12
$L_{Edd, NS}$ \citep{grimm:02}.

The hypothetical intermediate mass BHs, probably reaching
masses of $\sim 10^{2-5} M_{\odot}$, might be associated with
extremely high star formation rates (BHs merging in dense
stellar cluster etc.) and are expected to be significantly less
frequent than $\sim$stellar mass BHs. Therefore the
transition from the $\sim$stellar mass BH HMXB luminosity
function to intermediate mass BHs should be visible in the
cumulative luminosity function. Merging BHs are one possible
way of rapid growth of super-massive BHs that exist in
practically all galaxies. To accrete efficiently intermediate mass
BHs should form close binary systems with normal stars or be
in dense molecular clouds.

If the cut-off in the luminosity function, observed at $\sim$ few
$10^{40}$ erg s$^{-1}$ corresponds to the maximum possible luminosity of
$\sim$stellar mass BHs and if at $L>L_c$ the population of
hypothetical intermediate mass BHs emerges, it should lead to
a drastic change in the slope of the $L_X$--SFR relation at extreme
values of SFR \citep{gilfanov:02}. Therefore, observations of distant
star forming galaxies with very high SFR might be one of the best and
easiest ways to probe the population of intermediate mass black
holes.

\subsubsection{Three component luminosity function}

\begin{figure}
  \resizebox{\hsize}{!}{\includegraphics{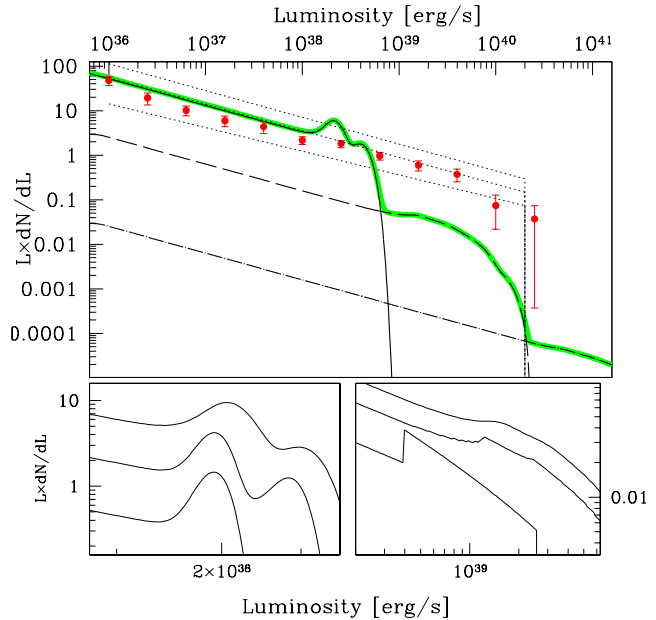}}
  \caption{The upper main figure shows the contributions of neutron
  stars (thin solid line), stellar (dashed line) and intermediate
  (dot-dashed line) mass BHs to the differential luminosity
  function. The thick grey solid line is the total differential
  luminosity function. For details of the parameters see discussion in
  the text. The figure in the lower left shows the luminosity around
  the Eddington limit for a NS. The luminosity functions
  shown include the simplest assumption that all systems with
  accretion rates above Eddington radiate at the Eddington luminosity
  (bottom), two effects allowing super-Eddington luminosities
  (middle), and additionally a 20 per cent uncertainty in the distance
  estimate (top). The curves are scaled for clarification. The figure
  in the lower right shows the luminosity around the Eddington limits
  for 3--20 M$_{\odot}$ BHs. The luminosity functions shown
  include no effect (bottom), and two effects allowing super-Eddington
  luminosities (middle) and additionally a 20 per cent uncertainty in the
  distance estimate (top). The dotted lines denote the uncertainty due
  to SFR of a factor of 2.}
\label{fig:comp_lf}
\end{figure}

In Fig. \ref{fig:comp_lf} we present the result of a simple picture of
what type of universal luminosity function a very simple model of HMXB
population synthesis could produce. This picture is obviously
oversimplified but we present it here to show that the simple picture
cannot reproduce the smooth luminosity function we get from CHANDRA
observations of star forming galaxies.

The initial set-up includes parameterisation of the mass
distributions of NSs and BHs, the distribution of
mass transfer rates in binary systems, and a prescription for the
conversion of mass transfer rates to X-ray luminosities.

The probability distribution of NS masses was chosen to be a
Gaussian distribution with a mean of 1.4 M$_{\odot}$ and a $\sigma$ of
0.2 M$_{\odot}$. The mass distribution of BHs was chosen to be
a power law with a slope of 1.1. These numbers are similar to
results of theoretical computations performed by \citet{fryer:01}. The
mass distribution for BHs is bimodal, for stellar mass black
holes it ranges from 3--20 M$_{\odot}$, and secondly, we include
intermediate mass BHs ranging from $10^{2}-10^{5}$
M$_{\odot}$. We made the simple assumption that their mass
distribution has the same slope as for stellar mass BHs.

Normalisations for the probability distributions were chosen such
that the number of stellar mass BHs is a factor of 20 smaller
than the number of NSs. This is roughly the ratio observed
for HMXBs in our Galaxy \citep{portegieszwart:98,iben:95,grimm:02}.
However the ratio of stars with $M> 25 M_{\odot}$, BH
progenitors, to stars with $25 M_{\odot} > M> 8 M_{\odot}$, NS
progenitors, is close to 1/2 according to the Salpeter IMF. Therefore
in principle the stellar mass BH curve in Fig. \ref{fig:comp_lf} might
be much closer to the NS curve. The number of intermediate mass BHs
is assumed to be a factor of 100 less than the number of stellar mass
BHs in HMXBs.

The probability distribution of mass transfer rates in binary systems
is set to be a power law with a slope of -1.6, reproducing the
observed luminosity function of HMXBs assuming a linear relation
between luminosity and mass accretion rate. The limits are 0.1 to
$10^{7}$ in units of $10^{16}$ g s$^{-1}$. Mass transfer was assumed to be
conservative over the whole range, i.e. no mass is lost from the
system except for super-Eddington sources and wind accretion. The
formulae for conversion of mass accretion rate to X-ray luminosity
are
\begin{eqnarray}
L = \eta \cdot \dot{M}_{accretion} \cdot c^2,
\end{eqnarray}
where $\eta = 0.1$ for BHs and $\eta = 0.15$ for NSs. The mass loss rate
from the normal star has no strict limit, however the X-ray luminosity
reaches a maximum at the Eddington luminosity and {\it objects with
much higher mass accretion rate will end up at the Eddington
luminosity introducing a peak in the luminosity function}.

For illustration we present two sub-figures in Fig. \ref{fig:comp_lf}
to show the evolution from sharp features to a smoother curve with the
introduction of smearing effects on the luminosity which is shown in
the main part of the figure.
The first effect is He-accretion when the HMXB is fed by a
helium rich star which we take to be the case in about 10 per cent of the
sources. Secondly, in the case of BHs a quasi-flat accretion
disk with an electron scattering atmosphere
\citep{sobolev:49,chandra:50} radiates according to $(1+2.08\cdot
\cos(i))\cos(i)$ where $i$ is the inclination angle, producing
2.6 times higher flux in the direction perpendicular to the disk plane
than average \citep{shakura:73}. \citet{sunyaev:85} confirmed that
this ratio is similar or higher for radiation comptonized in the
accretion disk. For slim disks \citep{paczynski:80} this ratio should
be even higher. Moreover to demonstrate the influence of distance
uncertainties we assumed a variation in distances of 20 per cent. All these
effects together give a considerably smoother curve and permit up to 6
times higher luminosities.

These are only the most simple effects that permit to surpass the
Eddington limit. Of course other more sophisticated models like jet
emission \citep{koerding:02} or beamed emission \citep{king:01} or
models taking into account strong magnetic fields in X-ray pulsars
\citep{basko:76} also can be employed to explain the observed
luminosity function.

\subsubsection{Wind driven accreting systems}
Our experience with HMXBs in our Galaxy and LMC shows that in many
sources accretion happens via capture from a strong stellar wind (Cen
X-3, Cyg X-1, 4U 1700+37, 4U 0900-40, and possibly SMC X-1, LMC X-1
and LMC X-4) As we see the majority of Galactic HMXBs are fed by
stellar wind accretion. There is a very important difference between
wind accretion onto NSs and BHs. The capture radius, $r_{capture} =
\frac{2GM}{v_{0}^{2}}$, is proportional to the mass of the accreting
object and therefore in similar systems BHs should have $M^2$ times
larger accretion rates than NSs for the same wind parameters. The
dependence of the Roche geometry on the mass ratio make the dependence
on M$_{BH}$ a little weaker.
\begin{equation}
  \dot{M}_{capture} \propto \dot{M}_{wind} \cdot
  (\frac{M_{BH}}{M_{NS}})^{\beta},
\end{equation}
where $\beta$ is between 1.5 and 2. This reason might increase the
relative BH contribution to the luminosity function in star forming
galaxies. It is important that
\begin{equation}
  \frac{\dot{M}_{capture}\cdot \eta \cdot c^2}{L_{Edd}} \propto
  \dot{M}_{wind} \cdot M_{BH}^{\beta-1}.
\end{equation}
For $\beta > 1$ it is preferable for BHs to have higher luminosities
than for NSs.

\subsubsection{Comparison of simulated and observed luminosity function}
The discrepancy between the observed luminosity function in the right
panel of Fig. \ref{fig:all_l} and our simple model in
Fig. \ref{fig:comp_lf} is obvious. We do not see features in the
observed differential luminosity function in the vicinity of $L_{Edd}$
for NSs, neither a peak $\frac{\Delta L}{L} \sim O(2)$ nor a sharp
decline at $L > L_{Edd}$ as in the model luminosity
function. Furthermore our model luminosity function lacks
sources in the luminosity range $10^{39}-10^{40}$ erg s$^{-1}$. It seems we
should assume that accreting stellar mass BHs in star forming regions
are more abundant than in the Milky Way.

It is important to note that having all our corrections we are
getting objects close to the limit of maximum luminosity of the
observed luminosity functions.

In Fig. \ref{fig:comp_lf} is plotted the total accretion
luminosity whereas CHANDRA observes only in the range from 1--10
keV. However X-ray pulsars emit the bulk of their luminosity in the
range from 20--40 keV. This effect may further decrease the importance
of the peak at $2\cdot 10^{38}$ erg s$^{-1}$. Since in elliptical galaxies
old X-ray binaries with weak magnetic fields, thus having much softer
spectra than X-ray pulsars, should dominate the population one should
expect the importance of the peak to be larger in ellipticals.

Our simple analysis demonstrates how difficult it is to construct a
very smooth luminosity function with the same slope over a broad
luminosity range, $10^{35}-10^{40}$ erg s$^{-1}$, and without sharp features
in the vicinity of Eddington luminosities. Because so many different
processes are involved in different parts of this huge luminosity
range. Our universal luminosity function based on CHANDRA, ASCA and
RXTE data has no strong features. The
absence of features around the Eddington luminosity for NSs should be
explained but it is also necessary to explain the absence of the
abrupt change in the luminosity function at higher luminosities when
less numerous BHs dominate the luminosity function.

The most obvious shortcomings of this naive model are the
mass distributions of BHs and NSs, the
normalisations for BHs, especially for intermediate mass BHs,
and the assumptions of conservative mass transfer and that all
super-Eddington sources radiate at Eddington luminosity in X-rays. It
is also very difficult to assume that intermediate mass BHs
form a continuous mass function with stellar mass BHs without
a strong break around 20--50 M$_{\odot}$. They should have their own
luminosity function with different normalisation and slope. Another
problem is connected with the formation of binaries with normal stars
feeding intermediate mass BHs and making them bright X-ray
sources. The observation of HMXBs in other galaxies will allow to put
constraints on the combination of these parameters.

The main concern with with the existence of a featureless universal
luminosity function (ULF) is connected with the interpretation of the
following experimental facts:
\begin{itemize}

\item RXTE/ASM, ASCA and CHANDRA give us information about the low
  luminosity part of the ULF ($L_X \la 10^{38}$ erg s$^{-1}$) based on the
  Milky Way, SMC and NGC 1569.

\item CHANDRA data on the other galaxies in Table \ref{tab:gal1} give
  information about the high luminosity part of the ULF ($L_X \ga
  10^{38}$ erg s$^{-1}$).

\item UV, FIR and radio methods of SFR determination in both local and
  more distant samples of galaxies have significant systematic
  uncertainties, see Table \ref{tab:flux}.
\end{itemize}
To resolve these uncertainties arising very close to the Eddington
luminosity for a NS we need to additional data permitting to get the
slope of the luminosity function in Antennae-type galaxies at
luminosities significantly below $10^{38}$ erg s$^{-1}$. Furthermore we need
to increase the sample of nearby galaxies where we can extend the
luminosity function well above $10^{38}$ erg s$^{-1}$. Only this will give
full confidence that there is no change in the normalisation in the
ULF near $10^{38}$ erg s$^{-1}$.

\subsection{Further astrophysically important information}

The good correlation between SFR and total X-ray luminosity due
to HMXBs and the total number of HMXBs can obviously become a powerful
and independent way to measure SFR in distant galaxies. In addition,
this correlation is providing us with further astrophysically
important information:
\begin{itemize}

\item These data are showing that NSs and BHs are
  produced in star forming regions very efficiently and in very short
  time, confirming the main predictions of stellar evolution. 

\item The luminosity function of HMXBs does not seem to depend
  strongly on the trigger of the star formation event which might be
  completely different for the Milky Way and e.g. the Antennae where
  it is the result of tidal interaction of two galaxies. 

\item The good agreement of the X-ray luminosity -- SFR relation of
  HDF galaxies with the theoretical prediction proves that the HMXB
  formation scenario at high redshifts does not differ significantly
  from nearby HMXB formation. 

\item The luminosity function provides information that neutron
  stars and BHs have a similar distribution of accretion rates
  in all galaxies of the sample available for study today.  

\item The luminosity function of HMXBs does not seem to depend
  strongly on the chemical abundances in the host galaxy.

\item The existence of well separated X-ray sources is a way to look
  for small satellites of massive galaxies, like SMC.
\end{itemize}

The integral X-ray luminosity and X-ray source counts are
{\it unique sources of information on binaries in distant
galaxies}. Other methods of investigation of SFR (UV, IR, radio) rely
on the luminosity distribution and number of the brightest stars,
without a significant dependence on the amount of binaries in a high
mass star population. On the other hand the existence of an observed
population of HMXBs in another galaxy is possible only in the case if
there are conditions for formation of close binaries with certain mass
loss from a normal companion and efficient capture of out-flowing
stellar wind or Roche lobe overflow by an accreting object. Detailed
observations of X-ray sources in our own Galaxy have shown how small
the allowed parameter space is -- this is the reason why the number of
X-ray sources in the Galaxy is so small \citep{illarionov:75} in
comparison with the total number of NSs and BHs and the total number
of O and B stars. Therefore:
\begin{itemize}
\item The existence of a universal luminosity function of
  HMXBs proves that the formation of close massive X-ray binaries and
  their distribution on mass ratio, separation and mass exchange rate
  is similar in all regions of active star formation up to redshifts
  z$\sim$1.
\end{itemize}

\section{Conclusion}

Based on CHANDRA and ASCA observations of nearby star forming galaxies
and RXTE/ASM, ASCA, and MIR-KVANT/TTM data on our Galaxy and the
Magellanic Clouds we studied the relation between star formation and
the population of high mass X-ray binaries. Within the accuracy and
completeness of the data available at present, we conclude that:

\begin{enumerate}
\renewcommand{\theenumi}{(\arabic{enumi})}

\item The data are broadly consistent with the assumption that in a
  wide range of star formation rates the luminosity distribution of
  HMXBs in a galaxy can be approximately described by a universal
  luminosity function, whose normalisation is proportional to the SFR
  (Fig. \ref{fig:lfs}, \ref{fig:comp_l}, \ref{fig:all_l}). Although the
  accuracy of this approximation is yet to be determined based on a
  larger galaxy sample and deeper observations, we conclude from the
  rather limited sample available, that it might be of the order of
  $\sim$50 per cent or better.

  In differential form the universal luminosity function can be
  approximated as a power law with a cut-off at $L_c\sim 2\cdot
  10^{40}$ erg s$^{-1}$:
\begin{equation}
\frac{dN}{dL_{38}}= (3.3^{+1.1}_{-0.8})\cdot
                  SFR \cdot L_{38}^{-1.61\pm0.12}
  \text{~~for } L<L_c,
\end{equation}
  where SFR is measured in units of M$_{\odot}$ yr$^{-1}$ and
  $L_{38}=L/10^{38}$ erg s$^{-1}$. In cumulative form it is
  correspondingly:
\begin{equation}
N(>L_{38})= (5.4^{+2.1}_{-1.7}) \cdot SFR \cdot (L_{38}^{-0.61 \pm
  0.12} -  210^{-0.61 \pm 0.12}).
\end{equation}

  Although more subtle effects can not presently be excluded (and are
  likely to exist), we did not find strong non-linear dependences of
  the HMXB luminosity function on SFR. We neither found strong
  dependences of the HMXB luminosity function on other parameters of
  the host galaxy, such as metalicity or star formation trigger. 

\item Both the number and total luminosity of HMXBs in a galaxy are
  directly related to the  star formation rate and can be used as an
  independent SFR indicator.

\item The total number of HMXBs is directly proportional to SFR
  (Fig. \ref{fig:numsfr}):
\begin{equation}
  \text{SFR [M$_{\odot}$ yr$^{-1}$]} = \frac{N(L>2 \cdot
  10^{38}\text{erg s$^{-1}$})}{2.9}.
\label{eq:sfr_n}
\end{equation}

\item The dependence of the total X-ray luminosity of a galaxy due to
  HMXBs on SFR has a break at SFR $\approx$ 4.5 M$_{\odot}$ yr$^{-1}$ for $M >
  8$ M$_{\odot}$.

  At sufficiently high values of star formation rate, SFR$\ga 4.5$
  M$_{\odot}$ yr$^{-1}$ ($L_{\rm 2-10~keV}\ga 3\cdot 10^{40}$ erg s$^{-1}$
  correspondingly) the X-ray luminosity of a galaxy due to HMXBs is
  directly proportional to SFR  (Fig.\ref{fig:l-sfr}):
\begin{equation}
{\rm SFR} [{\rm M}_{\odot} {\rm yr}^{-1}] = \frac{L_{\rm 2-10~keV}}{6.7\cdot
10^{39}{\rm erg~s}^{-1}}
\label{eq:sfr_lx}
\end{equation}
  At lower values of the star formation rate,  SFR$\la 4.5$
  M$_{\odot}$ yr$^{-1}$ ($L_{\rm 2-10~keV}\la 3\cdot 10^{40}$ erg
  s$^{-1}$), the $L_X-SFR$ relation is non-linear:
  (Fig.\ref{fig:l-sfr}):
\begin{equation}
{\rm SFR} [{\rm M}_{\odot} {\rm yr}^{-1}] = \left(\frac{L_{\rm 2-10~keV}}{2.6\cdot
10^{39}{\rm erg~s}^{-1}}\right)^{0.6}
\label{eq:sfr_lx1}
\end{equation}

  The non-linear $L_X-SFR$ dependence in the low SFR limit is {\it not}
  related to non-linear SFR dependent effects in the population of
  HMXB sources. It is rather caused by non-Gaussianity of the
  probability distribution of the integrated luminosity of a
  population of discrete sources. We will give this a more
  detailed and rigorous treatment in a forthcoming paper
  \citep{gilfanov:02}.

\item Based on the data of CHANDRA observations of the Hubble Deep
  Field North we showed, that the relation (\ref{eq:sfr_lx})  between
  the SFR and the X-ray luminosity of a galaxy due to HMXBs holds for
  distant star forming galaxies with redshifts as high as $z=1.2$
  (Fig. \ref{fig:l-sfr}).

\item The good agreement of high redshift observations with
  theoretical predictions and the fact that X-ray observations
  exclusively rely on the binary nature of the sources is evidence
  that not only the amount of star formation at redshifts up to
  $\sim$1 can be easily obtained from the above relations but also
  that the HMXB formation scenario is very similar at least up to this
  redshift.

\item The entire existence of the linear regime in the $L_X$--SFR
  relation is a direct consequence of the existence of a cut-off in
  the luminosity function. The position of the break in the $L_X-SFR$
  relation depends on the cut-off luminosity $L_c$ in the luminosity
  function of HMXB as SFR$_{\rm break}\propto L_c^{\alpha-1}$, where
  $\alpha$ is the differential slope of the luminosity function.
  Combined with the slope of the $L_X$--SFR relation in the low SFR
  regime (Eq.(\ref{eq:low_sfr})) this opens a possibility to constrain
  the parameters of the luminosity function studying the $L_X$--SFR
  relation alone, without actually constructing the luminosity
  functions, e.g. in distant unresolved galaxies.

  Agreement of the predicted $L_X-SFR$ relation with the data both
  in high and low SFR regime (Fig.\ref{fig:l-sfr}) gives an
  independent evidence of the existence of a cut-off in the luminosity
  function of HMXBs at $L_c\sim {\rm several} \times 10^{40}$ erg s$^{-1}$.
  It also indicates that $L_X-SFR$ data, including the high redshift
  galaxies from Hubble Deep Field North, are consistent with the HMXB
  luminosity function parameters, derived from significantly fewer
  galaxies, than plotted in Fig.\ref{fig:l-sfr}.

\end{enumerate}

\section*{acknowledgments}
We want to thank Jarle Brinchmann for helpful discussion about optical
properties of starburst galaxies and providing of data on the HDF
galaxies.

\addcontentsline{toc}{chapter}{Literaturverzeichnis}
   \bibliographystyle{mn2e}
   \bibliography{mc476}

\begin{thebibliography}{}

\bibitem[\protect\citeauthoryear{{Armus}, {Heckman} \& {Miley}}{{Armus}
  et~al.}{1990}]{armus:90}
{Armus} L.,  {Heckman} T.~M.,    {Miley} G.~K.,  1990, \apj, 364, 471

\bibitem[\protect\citeauthoryear{{Awaki}, {Matsumoto} \& {Tomida}}{{Awaki}
  et~al.}{2002}]{awaki:02}
{Awaki} H.,  {Matsumoto} H.,    {Tomida} H.,  2002, \apj, 567, 892

\bibitem[\protect\citeauthoryear{{Bahcall}}{{Bahcall}}{1983}]{bahcall:83}
{Bahcall} J.~N.,  1983, \apj, 267, 52

\bibitem[\protect\citeauthoryear{{Basko} \& {Sunyaev}}{{Basko} \&
  {Sunyaev}}{1976}]{basko:76}
{Basko} M.~M.,  {Sunyaev} R.~A.,  1976, \mnras, 175, 395

\bibitem[\protect\citeauthoryear{{Bell} \& {Kennicutt}}{{Bell} \&
  {Kennicutt}}{2001}]{bell:01}
{Bell} E.~F.,  {Kennicutt} R.~C.,  2001, \apj, 548, 681

\bibitem[\protect\citeauthoryear{{Bolton}}{{Bolton}}{1972}]{bolton:72}
{Bolton} C.~T.,  1972, Nature Physical Science, 240, 124+

\bibitem[\protect\citeauthoryear{{Brandt}, {Alexander}, {Hornschemeier},
  {Garmire}, {Schneider}, {Barger}, {Bauer}, {Broos}, {Cowie}, {Townsley},
  {Burrows}, {Chartas}, {Feigelson}, {Griffiths}, {Nousek} \&
  {Sargent}}{{Brandt} et~al.}{2001}]{brandt:01}
{Brandt} W.~N.,  {Alexander} D.~M.,  {Hornschemeier} A.~E.,  {Garmire} G.~P.,
  {Schneider} D.~P.,  {Barger} A.~J.,  {Bauer} F.~E.,  {Broos} P.~S.,  {Cowie}
  L.~L.,  {Townsley} L.~K.,  {Burrows} D.~N.,  {Chartas} G.,  {Feigelson}
  E.~D.,  {Griffiths} R.~E.,  {Nousek} J.~A.,    {Sargent} W.~L.~W.,  2001,
  \aj, 122, 2810

\bibitem[\protect\citeauthoryear{{Brinchmann} \& {Ellis}}{{Brinchmann} \&
  {Ellis}}{2000}]{brinchmann:00}
{Brinchmann} J.,  {Ellis} R.~S.,  2000, \apjl, 536, L77

\bibitem[\protect\citeauthoryear{{Buat}, {Boselli}, {Gavazzi} \&
  {Bonfanti}}{{Buat} et~al.}{2002}]{buat:02}
{Buat} V.,  {Boselli} A.,  {Gavazzi} G.,    {Bonfanti} C.,  2002, \aap, 383,
  801

\bibitem[\protect\citeauthoryear{{Chandrasekhar}}{{Chandrasekhar}}{1950}]{chan%
dra:50}
{Chandrasekhar} S.,  1950, {Radiative transfer.}.
Oxford, Clarendon Press, 1950.

\bibitem[\protect\citeauthoryear{{Cohen}, {Hogg}, {Blandford}, {Cowie}, {Hu},
  {Songaila}, {Shopbell} \& {Richberg}}{{Cohen} et~al.}{2000}]{cohen:00}
{Cohen} J.~G.,  {Hogg} D.~W.,  {Blandford} R.,  {Cowie} L.~L.,  {Hu} E.,
  {Songaila} A.,  {Shopbell} P.,    {Richberg} K.,  2000, \apj, 538, 29

\bibitem[\protect\citeauthoryear{{Condon}}{{Condon}}{1992}]{condon:92}
{Condon} J.~J.,  1992, \araa, 30, 575

\bibitem[\protect\citeauthoryear{{Condon}, {Helou}, {Sanders} \&
  {Soifer}}{{Condon} et~al.}{1990}]{condon:90}
{Condon} J.~J.,  {Helou} G.,  {Sanders} D.~B.,    {Soifer} B.~T.,  1990, \apjs,
  73, 359

\bibitem[\protect\citeauthoryear{{David}, {Jones} \& {Forman}}{{David}
  et~al.}{1992}]{david:92}
{David} L.~P.,  {Jones} C.,    {Forman} W.,  1992, \apj, 388, 82

\bibitem[\protect\citeauthoryear{{de Vaucouleurs}, {de Vaucouleurs}, {Corwin},
  {Buta}, {Paturel} \& {Fouque}}{{de Vaucouleurs}
  et~al.}{1991}]{devaucouleurs:91}
{de Vaucouleurs} G.,  {de Vaucouleurs} A.,  {Corwin} H.~G.,  {Buta} R.~J.,
  {Paturel} G.,    {Fouque} P.,  1991, {Third Reference Catalogue of Bright
  Galaxies}.
Volume 1-3, XII, 2069 pp.~7 figs..~ Springer-Verlag Berlin Heidelberg New York

\bibitem[\protect\citeauthoryear{{Eneev}, {Kozlov} \& {Sunyaev}}{{Eneev}
  et~al.}{1973}]{eneev:73}
{Eneev} T.~M.,  {Kozlov} N.~N.,    {Sunyaev} R.~A.,  1973, \aap, 22, 41+

\bibitem[\protect\citeauthoryear{Eracleous, Shields, Chartas \&
  Moran}{Eracleous et~al.}{2002}]{eracleous:02}
Eracleous M.,  Shields J.~C.,  Chartas G.,    Moran E.~C.,  2002, \apj, 565,
  108

\bibitem[\protect\citeauthoryear{Fabbiano}{Fabbiano}{1994}]{fabbiano:94}
Fabbiano G.,  1994, X-ray binaries.
Cambridge University Press, p.~390

\bibitem[\protect\citeauthoryear{{Fabbiano}, {Gioia} \&
  {Trinchieri}}{{Fabbiano} et~al.}{1988}]{fabbiano:88}
{Fabbiano} G.,  {Gioia} I.~M.,    {Trinchieri} G.,  1988, \apj, 324, 749

\bibitem[\protect\citeauthoryear{{Feitzinger}}{{Feitzinger}}{1980}]{feitzinger%
:80}
{Feitzinger} J.~V.,  1980, Space Science Reviews, 27, 35

\bibitem[\protect\citeauthoryear{{Fryer} \& {Kalogera}}{{Fryer} \&
  {Kalogera}}{2001}]{fryer:01}
{Fryer} C.~L.,  {Kalogera} V.,  2001, \apj, 554, 548

\bibitem[\protect\citeauthoryear{{Galletta} \& {Recillas-Cruz}}{{Galletta} \&
  {Recillas-Cruz}}{1982}]{galletta:82}
{Galletta} G.,  {Recillas-Cruz} E.,  1982, \aap, 112, 361

\bibitem[\protect\citeauthoryear{{Georgakakis}, {Forbes} \&
  {Norris}}{{Georgakakis} et~al.}{2000}]{georgakakis:00}
{Georgakakis} A.,  {Forbes} D.~A.,    {Norris} R.~P.,  2000, \mnras, 318, 124

\bibitem[\protect\citeauthoryear{{Ghosh} \& {White}}{{Ghosh} \&
  {White}}{2001}]{ghosh:01}
{Ghosh} P.,  {White} N.~E.,  2001, \apjl, 559, L97

\bibitem[\protect\citeauthoryear{Gilfanov, Grimm \& Sunyaev}{Gilfanov
  et~al.}{2002}]{gilfanov:02}
Gilfanov M.,  Grimm H.-J.,    Sunyaev R.,  2002, Collective luminosity of a
  population of discrete sources: Astrophysical implications, in preparation

\bibitem[\protect\citeauthoryear{{Gonzalez Delgado} \& {Perez}}{{Gonzalez
  Delgado} \& {Perez}}{1996}]{gonzalez:96}
{Gonzalez Delgado} R.~M.,  {Perez} E.,  1996, \mnras, 281, 1105

\bibitem[\protect\citeauthoryear{{Greiner}, {Cuby} \& {McCaughrean}}{{Greiner}
  et~al.}{2001}]{greiner:01}
{Greiner} J.,  {Cuby} J.~G.,    {McCaughrean} M.~J.,  2001, \nat, 414, 522

\bibitem[\protect\citeauthoryear{Griffiths, Ptak, Feigelson, Garmire, Townsley,
  Brandt, Sambruna \& Bregman}{Griffiths et~al.}{2000}]{griffiths:00}
Griffiths R.,  Ptak A.,  Feigelson E.,  Garmire G.,  Townsley L.,  Brandt W.,
  Sambruna R.,    Bregman J.,  2000, Science, 290, 1325

\bibitem[\protect\citeauthoryear{{Griffiths} \& {Padovani}}{{Griffiths} \&
  {Padovani}}{1990}]{griffiths:90}
{Griffiths} R.~E.,  {Padovani} P.,  1990, \apj, 360, 483

\bibitem[\protect\citeauthoryear{{Grimm}, {Gilfanov} \& {Sunyaev}}{{Grimm}
  et~al.}{2002}]{grimm:02}
{Grimm} H.-J.,  {Gilfanov} M.,    {Sunyaev} R.,  2002, \aap, 391, 923

\bibitem[\protect\citeauthoryear{{Holtzman}, {Gallagher}, {Cole}, {Mould},
  {Grillmair}, {Ballester}, {Burrows}, {Clarke}, {Crisp}, {Evans}, {Griffiths},
  {Hester}, {Hoessel}, {Scowen}, {Stapelfeldt}, {Trauger} \&
  {Watson}}{{Holtzman} et~al.}{1999}]{holtzman:99}
{Holtzman} J.~A.,  {Gallagher} J.~S.,  {Cole} A.~A.,  {Mould} J.~R.,
  {Grillmair} C.~J.,  {Ballester} G.~E.,  {Burrows} C.~J.,  {Clarke} J.~T.,
  {Crisp} D.,  {Evans} R.~W.,  {Griffiths} R.~E.,  {Hester} J.~J.,  {Hoessel}
  J.~G.,  {Scowen} P.~A.,  {Stapelfeldt} K.~R.,  {Trauger} J.~T.,    {Watson}
  A.~M.,  1999, \aj, 118, 2262

\bibitem[\protect\citeauthoryear{{Hoopes}, {Walterbos} \& {Bothun}}{{Hoopes}
  et~al.}{2001}]{hoopes:01}
{Hoopes} C.~G.,  {Walterbos} R.~.~M.,    {Bothun} G.~D.,  2001, \apj, 559, 878

\bibitem[\protect\citeauthoryear{{Huchtmeier} \& {Richter}}{{Huchtmeier} \&
  {Richter}}{1988}]{huchtmeier:88}
{Huchtmeier} W.~K.,  {Richter} O.-G.,  1988, \aap, 203, 237

\bibitem[\protect\citeauthoryear{{Iben}, {Tutukov} \& {Yungelson}}{{Iben}
  et~al.}{1995}]{iben:95}
{Iben} I.~J.,  {Tutukov} A.~V.,    {Yungelson} L.~R.,  1995, \apjs, 100, 217

\bibitem[\protect\citeauthoryear{{Illarionov} \& {Sunyaev}}{{Illarionov} \&
  {Sunyaev}}{1975}]{illarionov:75}
{Illarionov} A.~F.,  {Sunyaev} R.~A.,  1975, \aap, 39, 185

\bibitem[\protect\citeauthoryear{{Israel}}{{Israel}}{1988}]{israel:88}
{Israel} F.~P.,  1988, \aap, 194, 24

\bibitem[\protect\citeauthoryear{{K{\" o}rding}, {Falcke} \& {Markoff}}{{K{\"
  o}rding} et~al.}{2002}]{koerding:02}
{K{\" o}rding} E.,  {Falcke} H.,    {Markoff} S.,  2002, \aap, 382, L13

\bibitem[\protect\citeauthoryear{{Kaaret}}{{Kaaret}}{2001}]{kaaret:01}
{Kaaret} P.,  2001, \apj, 560, 715

\bibitem[\protect\citeauthoryear{{Kalogera}, {Narayan}, {Spergel} \&
  {Taylor}}{{Kalogera} et~al.}{2001}]{kalogera:01}
{Kalogera} V.,  {Narayan} R.,  {Spergel} D.~N.,    {Taylor} J.~H.,  2001, \apj,
  556, 340

\bibitem[\protect\citeauthoryear{{Karachentsev}, {Drozdovsky}, {Kajsin},
  {Takalo}, {Heinamaki} \& {Valtonen}}{{Karachentsev}
  et~al.}{1997}]{karachentsev:97}
{Karachentsev} I.,  {Drozdovsky} I.,  {Kajsin} S.,  {Takalo} L.~O.,
  {Heinamaki} P.,    {Valtonen} M.,  1997, \aaps, 124, 559

\bibitem[\protect\citeauthoryear{{Kennicutt}}{{Kennicutt}}{1998}]{kennicutt:98}
{Kennicutt} R.~C.,  1998, \araa, 36, 189

\bibitem[\protect\citeauthoryear{{Kennicutt}, {Tamblyn} \&
  {Congdon}}{{Kennicutt} et~al.}{1994}]{kennicutt:94}
{Kennicutt} R.~C.,  {Tamblyn} P.,    {Congdon} C.~E.,  1994, \apj, 435, 22

\bibitem[\protect\citeauthoryear{{Kilgard}, {Kaaret}, {Krauss}, {McDowell},
  {Prestwich}, {Raley} \& {Zezas}}{{Kilgard} et~al.}{2001}]{kilgard:01}
{Kilgard} R.~E.,  {Kaaret} P.~E.,  {Krauss} M.~I.,  {McDowell} J.~C.,
  {Prestwich} A.~H.,  {Raley} M.~T.,    {Zezas} A.,  2001, American
  Astronomical Society Meeting

\bibitem[\protect\citeauthoryear{King, Davies, Ward, Fabbiano \& Elvis}{King
  et~al.}{2001}]{king:01}
King A.,  Davies M.,  Ward M.,  Fabbiano G.,    Elvis M.,  2001, ApJ, 552, L109

\bibitem[\protect\citeauthoryear{{Kuno} \& {Nakai}}{{Kuno} \&
  {Nakai}}{1997}]{kuno:97}
{Kuno} N.,  {Nakai} N.,  1997, \pasj, 49, 279

\bibitem[\protect\citeauthoryear{{L{\' i}pari}, {D{\' i}az}, {Taniguchi},
  {Terlevich}, {Dottori} \& {Carranza}}{{L{\' i}pari} et~al.}{2000}]{lipari:00}
{L{\' i}pari} S.,  {D{\' i}az} R.,  {Taniguchi} Y.,  {Terlevich} R.,  {Dottori}
  H.,    {Carranza} G.,  2000, \aj, 120, 645

\bibitem[\protect\citeauthoryear{{Lehnert} \& {Heckman}}{{Lehnert} \&
  {Heckman}}{1996}]{lehnert:96}
{Lehnert} M.~D.,  {Heckman} T.~M.,  1996, \apj, 472, 546+

\bibitem[\protect\citeauthoryear{{Lira}, {Ward}, {Zezas}, {Alonso-Herrero} \&
  {Ueno}}{{Lira} et~al.}{2002}]{lira:02}
{Lira} P.,  {Ward} M.,  {Zezas} A.,  {Alonso-Herrero} A.,    {Ueno} S.,  2002,
  \mnras, 330, 259

\bibitem[\protect\citeauthoryear{{Liu} \& {Kennicutt}}{{Liu} \&
  {Kennicutt}}{1995}]{liu:95}
{Liu} C.~T.,  {Kennicutt} R.~C.,  1995, \apj, 450, 547+

\bibitem[\protect\citeauthoryear{{Lyutyi}, {Syunyaev} \&
  {Cherepashchuk}}{{Lyutyi} et~al.}{1973}]{lyutyi:73}
{Lyutyi} V.~M.,  {Syunyaev} R.~A.,    {Cherepashchuk} A.~M.,  1973, Soviet
  Astronomy, 17, 1+

\bibitem[\protect\citeauthoryear{{Madau} \& {Pozzetti}}{{Madau} \&
  {Pozzetti}}{2000}]{madau:00}
{Madau} P.,  {Pozzetti} L.,  2000, \mnras, 312, L9

\bibitem[\protect\citeauthoryear{{Martin}, {Kobulnicky} \& {Heckman}}{{Martin}
  et~al.}{2002}]{martin:02}
{Martin} C.~L.,  {Kobulnicky} H.~A.,    {Heckman} T.~M.,  2002, \apj, 574, 663

\bibitem[\protect\citeauthoryear{{Moshir}, {Copan}, {Conrow}, {McCallon},
  {Hacking}, {Gregorich}, {Rohrbach}, {Melnyk}, {Rice} \& {Fullmer}}{{Moshir}
  et~al.}{1993}]{moshir:93}
{Moshir} M.,  {Copan} G.,  {Conrow} T.,  {McCallon} H.,  {Hacking} P.,
  {Gregorich} D.,  {Rohrbach} G.,  {Melnyk} M.,  {Rice} W.,    {Fullmer} L.,
  1993, VizieR Online Data Catalog, 2156, 0

\bibitem[\protect\citeauthoryear{{Negishi}, {Onaka}, {Chan} \&
  {Roellig}}{{Negishi} et~al.}{2001}]{negishi:01}
{Negishi} T.,  {Onaka} T.,  {Chan} K.-W.,    {Roellig} T.~L.,  2001, \aap, 375,
  566

\bibitem[\protect\citeauthoryear{{Oort}}{{Oort}}{1987}]{oort:87}
{Oort} M.~J.~A.,  1987, \aaps, 71, 221

\bibitem[\protect\citeauthoryear{Paczynsky \& Wiita}{Paczynsky \&
  Wiita}{1980}]{paczynski:80}
Paczynsky B.,  Wiita P.,  1980, A\&A, 88, 23

\bibitem[\protect\citeauthoryear{{Persic} \& {Salucci}}{{Persic} \&
  {Salucci}}{1988}]{persic:88}
{Persic} M.,  {Salucci} P.,  1988, \mnras, 234, 131

\bibitem[\protect\citeauthoryear{{Portegies Zwart} \& {Yungelson}}{{Portegies
  Zwart} \& {Yungelson}}{1998}]{portegieszwart:98}
{Portegies Zwart} S.~F.,  {Yungelson} L.~R.,  1998, \aap, 332, 173

\bibitem[\protect\citeauthoryear{{Ptak}, {Griffiths}, {White} \&
  {Ghosh}}{{Ptak} et~al.}{2001}]{ptak:01}
{Ptak} A.,  {Griffiths} R.,  {White} N.,    {Ghosh} P.,  2001, \apjl, 559, L91

\bibitem[\protect\citeauthoryear{Ranalli, Comastri \& Setti}{Ranalli
  et~al.}{2002}]{ranalli:02}
Ranalli P.,  Comastri A.,    Setti G.,  2002, The 2--10 keV luminosity as a
  star formation rate indicator, astro-ph/0202241

\bibitem[\protect\citeauthoryear{{Reakes}}{{Reakes}}{1980}]{reakes:80}
{Reakes} M.,  1980, \mnras, 192, 297

\bibitem[\protect\citeauthoryear{{Rephaeli} \& {Gruber}}{{Rephaeli} \&
  {Gruber}}{2002}]{rephaeli:02}
{Rephaeli} Y.,  {Gruber} D.,  2002, \aap, 389, 752

\bibitem[\protect\citeauthoryear{{Rice}, {Lonsdale}, {Soifer}, {Neugebauer},
  {Koplan}, {Lloyd}, {de Jong} \& {Habing}}{{Rice} et~al.}{1988}]{rice:88}
{Rice} W.,  {Lonsdale} C.~J.,  {Soifer} B.~T.,  {Neugebauer} G.,  {Koplan}
  E.~L.,  {Lloyd} L.~A.,  {de Jong} T.,    {Habing} H.~J.,  1988, \apjs, 68, 91

\bibitem[\protect\citeauthoryear{{Richards}}{{Richards}}{2000}]{richards:00}
{Richards} E.~A.,  2000, \apj, 533, 611

\bibitem[\protect\citeauthoryear{{Richards}, {Kellermann}, {Fomalont},
  {Windhorst} \& {Partridge}}{{Richards} et~al.}{1998}]{richards:98}
{Richards} E.~A.,  {Kellermann} K.~I.,  {Fomalont} E.~B.,  {Windhorst} R.~A.,
   {Partridge} R.~B.,  1998, \aj, 116, 1039

\bibitem[\protect\citeauthoryear{Roberts, Warwick, Ward \& Murray}{Roberts
  et~al.}{2002}]{roberts:02}
Roberts T.,  Warwick R.,  Ward M.,    Murray S.,  2002, A Chandra observation
  of the interacting pair of glaaxies NGC 4485/NGC 4490, astro-ph/0208196

\bibitem[\protect\citeauthoryear{Rosa-Gonzalez, Terlevich \&
  Terlevich}{Rosa-Gonzalez et~al.}{2002}]{rosa-gonzales:02}
Rosa-Gonzalez D.,  Terlevich E.,    Terlevich R.,  2002, An Empirical
  Calibration of Star Formation Rate Estimators, accepted by MNRAS

\bibitem[\protect\citeauthoryear{{Roussel}, {Sauvage}, {Vigroux} \&
  {Bosma}}{{Roussel} et~al.}{2001}]{roussel:01}
{Roussel} H.,  {Sauvage} M.,  {Vigroux} L.,    {Bosma} A.,  2001, \aap, 372,
  427

\bibitem[\protect\citeauthoryear{{Rownd} \& {Young}}{{Rownd} \&
  {Young}}{1999}]{rownd:99}
{Rownd} B.~K.,  {Young} J.~S.,  1999, \aj, 118, 670

\bibitem[\protect\citeauthoryear{{Sage}}{{Sage}}{1993}]{sage:93}
{Sage} L.~J.,  1993, \aap, 272, 123

\bibitem[\protect\citeauthoryear{{Sandage} \& {Tammann}}{{Sandage} \&
  {Tammann}}{1980}]{sandage:80}
{Sandage} A.,  {Tammann} G.~A.,  1980, {A revised Shapley-Ames Catalog of
  bright galaxies}.
Washington: Carnegie Institution, 1980, Preliminary version

\bibitem[\protect\citeauthoryear{{Schreier}, {Giacconi}, {Gursky}, {Kellogg} \&
  {Tananbaum}}{{Schreier} et~al.}{1972}]{schreier:72}
{Schreier} E.,  {Giacconi} R.,  {Gursky} H.,  {Kellogg} E.,    {Tananbaum} H.,
  1972, \apjl, 178, L71

\bibitem[\protect\citeauthoryear{Schurch, Roberts \& Warwick}{Schurch
  et~al.}{2002}]{schurch:02}
Schurch N.,  Roberts T.,    Warwick R.,  2002, High-resolution X-ray imaging
  and spectroscopy of the core of NGC 4945 with XMM-Newton and Chandra,
  astro-ph/0204361

\bibitem[\protect\citeauthoryear{Shakura \& Sunyaev}{Shakura \&
  Sunyaev}{1973}]{shakura:73}
Shakura N.,  Sunyaev R.,  1973, \aap, 24, 337

\bibitem[\protect\citeauthoryear{{Sharina}, {Karachentsev} \&
  {Tikhonov}}{{Sharina} et~al.}{1996}]{sharina:96}
{Sharina} M.~E.,  {Karachentsev} I.~D.,    {Tikhonov} N.~A.,  1996, \aaps, 119,
  499

\bibitem[\protect\citeauthoryear{{Smith} \& {Wilson}}{{Smith} \&
  {Wilson}}{2001}]{smith:01}
{Smith} D.~A.,  {Wilson} A.~S.,  2001, \apj, 557, 180

\bibitem[\protect\citeauthoryear{Sobolev}{Sobolev}{1949}]{sobolev:49}
Sobolev V.,  1949, Seria Matem. Nauk, 18, N1163

\bibitem[\protect\citeauthoryear{{Sofue}, {Reuter}, {Krause}, {Wielebinski} \&
  {Nakai}}{{Sofue} et~al.}{1992}]{sofue:92}
{Sofue} Y.,  {Reuter} H.-P.,  {Krause} M.,  {Wielebinski} R.,    {Nakai} N.,
  1992, \apj, 395, 126

\bibitem[\protect\citeauthoryear{{Soria} \& {Kong}}{{Soria} \&
  {Kong}}{2002}]{soria:02b}
{Soria} R.,  {Kong} A.~K.~H.,  2002, \apjl, 572, L33

\bibitem[\protect\citeauthoryear{{Soria} \& {Wu}}{{Soria} \&
  {Wu}}{2002}]{soria:02}
{Soria} R.,  {Wu} K.,  2002, \aap, 384, 99

\bibitem[\protect\citeauthoryear{{Spinoglio}, {Malkan}, {Rush}, {Carrasco} \&
  {Recillas-Cruz}}{{Spinoglio} et~al.}{1995}]{spinoglio:95}
{Spinoglio} L.,  {Malkan} M.~A.,  {Rush} B.,  {Carrasco} L.,    {Recillas-Cruz}
  E.,  1995, \apj, 453, 616

\bibitem[\protect\citeauthoryear{{Stern}, {Tozzi}, {Stanford}, {Rosati},
  {Holden}, {Eisenhardt}, {Elston}, {Wu}, {Connolly}, {Spinrad}, {Dawson},
  {Dey} \& {Chaffee}}{{Stern} et~al.}{2002}]{stern:02}
{Stern} D.,  {Tozzi} P.,  {Stanford} S.~A.,  {Rosati} P.,  {Holden} B.,
  {Eisenhardt} P.,  {Elston} R.,  {Wu} K.~L.,  {Connolly} A.,  {Spinrad} H.,
  {Dawson} S.,  {Dey} A.,    {Chaffee} F.~H.,  2002, \aj, 123, 2223

\bibitem[\protect\citeauthoryear{{Strickland}, {Colbert}, {Heckman}, {Weaver},
  {Dahlem} \& {Stevens}}{{Strickland} et~al.}{2001}]{strickland:01}
{Strickland} D.~K.,  {Colbert} E.~J.~M.,  {Heckman} T.~M.,  {Weaver} K.~A.,
  {Dahlem} M.,    {Stevens} I.~R.,  2001, \apj, 560, 707

\bibitem[\protect\citeauthoryear{{Sunyaev}, {Gilfanov}, {Churazov}, {Loznikov},
  {Yamburenko}, {Skinner}, {Patterson}, {Willmore}, {Emam}, {Brinkman},
  {Heise}, {Intzand} \& {Jager}}{{Sunyaev} et~al.}{1990}]{sunyaev:90}
{Sunyaev} R.,  {Gilfanov} M.,  {Churazov} E.,  {Loznikov} V.,  {Yamburenko} N.,
   {Skinner} G.~K.,  {Patterson} T.~G.,  {Willmore} A.~P.,  {Emam} O.,
  {Brinkman} A.~C.,  {Heise} J.,  {Intzand} J.,    {Jager} R.,  1990, Soviet
  Astronomy Letters, 16, 55+

\bibitem[\protect\citeauthoryear{{Sunyaev}, {Tinsley} \& {Meier}}{{Sunyaev}
  et~al.}{1978}]{sunyaev:78}
{Sunyaev} R.~A.,  {Tinsley} B.~M.,    {Meier} D.~L.,  1978, Comments on
  Astrophysics, 7, 183

\bibitem[\protect\citeauthoryear{{Sunyaev} \& {Titarchuk}}{{Sunyaev} \&
  {Titarchuk}}{1985}]{sunyaev:85}
{Sunyaev} R.~A.,  {Titarchuk} L.~G.,  1985, \aap, 143, 374

\bibitem[\protect\citeauthoryear{{Tananbaum}, {Gursky}, {Kellogg}, {Giacconi}
  \& {Jones}}{{Tananbaum} et~al.}{1972}]{tananbaum:72}
{Tananbaum} H.,  {Gursky} H.,  {Kellogg} E.,  {Giacconi} R.,    {Jones} C.,
  1972, \apjl, 177, L5

\bibitem[\protect\citeauthoryear{Terashima \& Wilson}{Terashima \&
  Wilson}{2002}]{terashima:02}
Terashima Y.,  Wilson A.,  2002, Ultraluminous X-ray sources in M51,
  astro-ph/0204321

\bibitem[\protect\citeauthoryear{{Thronson}, {Hunter}, {Casey}, {Harper} \&
  {Latter}}{{Thronson} et~al.}{1989}]{thronson:89}
{Thronson} H.~A.,  {Hunter} D.~A.,  {Casey} S.,  {Harper} D.~A.,    {Latter}
  W.~B.,  1989, \apj, 339, 803

\bibitem[\protect\citeauthoryear{{Toomre} \& {Toomre}}{{Toomre} \&
  {Toomre}}{1972}]{toomre:72}
{Toomre} A.,  {Toomre} J.,  1972, \apj, 178, 623

\bibitem[\protect\citeauthoryear{{Ueda}, {Ishisaki}, {Takahashi}, {Makishima}
  \& {Ohashi}}{{Ueda} et~al.}{2001}]{ueda:01}
{Ueda} Y.,  {Ishisaki} Y.,  {Takahashi} T.,  {Makishima} K.,    {Ohashi} T.,
  2001, \apjs, 133, 1

\bibitem[\protect\citeauthoryear{Verbunt \& van~den Heuvel}{Verbunt \& van~den
  Heuvel}{1994}]{verbunt:94}
Verbunt F.,  van~den Heuvel E.,  1994, X-ray binaries.
Cambridge University Press, p.~457

\bibitem[\protect\citeauthoryear{{Viallefond}, {Allen} \& {de
  Boer}}{{Viallefond} et~al.}{1980}]{viallefond:80}
{Viallefond} F.,  {Allen} R.~J.,    {de Boer} J.~A.,  1980, \aap, 82, 207

\bibitem[\protect\citeauthoryear{{Wilkinson} \& {Evans}}{{Wilkinson} \&
  {Evans}}{1999}]{wilkinson:99}
{Wilkinson} M.~I.,  {Evans} N.~W.,  1999, \mnras, 310, 645

\bibitem[\protect\citeauthoryear{{Yokogawa}, {Imanishi}, {Tsujimoto},
  {Nishiuchi}, {Koyama}, {Nagase} \& {Corbet}}{{Yokogawa}
  et~al.}{2000}]{yokogawa:00}
{Yokogawa} J.,  {Imanishi} K.,  {Tsujimoto} M.,  {Nishiuchi} M.,  {Koyama} K.,
  {Nagase} F.,    {Corbet} R.~H.~D.,  2000, \apjs, 128, 491

\bibitem[\protect\citeauthoryear{{Young}, {Allen}, {Kenney}, {Lesser} \&
  {Rownd}}{{Young} et~al.}{1996}]{young:96}
{Young} J.~S.,  {Allen} L.,  {Kenney} J.~D.~P.,  {Lesser} A.,    {Rownd} B.,
  1996, \aj, 112, 1903

\bibitem[\protect\citeauthoryear{Zezas, Fabbiano, Rots \& Murray}{Zezas
  et~al.}{2002}]{zezas:02}
Zezas A.,  Fabbiano G.,  Rots A.,    Murray S.,  2002, Chandra observations of
  "The Antennae" galaxies (NGC4038/39): II. Detection and analysis of galaxian
  X-ray sources, astro-ph/0203174

\end{thebibliography}
\clearpage
\end{document}